\documentclass[twocolumn,aps,prd,amsmath,showpacs,floatfix,reprint,eprint]{revtex4-2} 
\usepackage{amsmath}
\usepackage{mathtools}
\usepackage{siunitx}
\usepackage[nameinlink,capitalize]{cleveref}
\usepackage{comment}
\usepackage{textcomp}
\usepackage{subcaption}
\usepackage[inline]{enumitem}

\usepackage[dvipsnames]{xcolor}

\newenvironment{change}{%
  \color{red}%
}{%
}

\newcommand{\substitute}[2] {%
  \Bigg[ #1 \Bigg]_{#2}
}%

\date{\today}
\begin{document}
\title{Relativistic Langevin equation derived from a particle-bath Lagrangian}

\author{Aleksandr \surname{Petrosyan}}
\email{ap886@cantab.cam.ac.uk}
\affiliation{Cavendish Laboratory, University of Cambridge, JJ Thomson Avenue, CB3 0HE, Cambridge, U.K.}

\author{Alessio \surname{Zaccone}}
\email{az302@cam.ac.uk}
\affiliation{Cavendish Laboratory, University of Cambridge, JJ Thomson Avenue, CB3 0HE, Cambridge, U.K.}
\affiliation{Department of Physics ``A. Pontremoli'', University of Milan, via Celoria 16, 20133 Milan, Italy.}
\begin{abstract}

We show how a relativistic Langevin equation can be derived from a
  Lorentz-covariant version of the Caldeira-Leggett particle-bath
  Lagrangian. In one of its limits, we identify the obtained equation
  with the Langevin equation used in contemporary extensions of
  statistical mechanics to the near-light-speed motion of a tagged
  particle in non-relativistic dissipative fluids. The proposed
  framework provides a more rigorous and first-principles form of the
  weakly-relativistic and partially-relativistic Langevin equations often quoted or postulated as ansatz in previous
  works. We then refine the aforementioned results to obtain a generalized Langevin equation valid for the case of both fully-relativistic particle and bath, using an analytical approximation obtained from numerics where the Fourier modes of the bath are systematically replaced with covariant plane-wave forms with a length-scale relativistic correction that depends on the space-time trajectory in a parabolic way. We discuss the implications of the apparent breaking of space-time
  translation and parity invariance, showing that these effects are
  not necessarily in contradiction with the assumptions of statistical
  mechanics. The intrinsically non-Markovian character of the fully
  relativistic generalised Langevin equation derived here, and of the
  associated fluctuation-dissipation theorem, is also discussed.

\end{abstract}

\maketitle

\section{Introduction}\label{sec:introduction}

The Langevin equation is the cornerstone of modern nonequilibrium
statistical mechanics~\cite{zwanzig2001nonequilibrium,9780486815558},
with profound implications for numerical
computations~\cite{hu2020meanfield,Parrinello}. It is the basis upon
which one builds the diffusion
equations~\cite{LI2020105481,FA2020123334,CATTIAUX2019108288,Relat-langev-dyn-in-exp-media},
fluctuation-dissipation theorems~\cite{zwanzig2001nonequilibrium,Cui},
Brownian motion with state-dependent diffusion coefficient and its
path-integral formulation~\cite{Lubensky}, and instrumental in areas
overlapping with other fundamental theories of physics, such as the
question of quantum decoherence~\cite{Hollowood_2017,Breuer}. Though
it may appear to be a simple reformulation of Newton's second law that
accounts for thermal noise and frictional damping, it encapsulates the
modern view of the structure of matter and the dynamics of Brownian
motion. For example, the Langevin equation summarizes our fundamental
understanding of molecular motion in liquids as systems made up of
stochastically moving molecules, on which short-range collisions and
interactions with other surrounding molecules impart both a stochastic
excitation and a viscous damping~\cite{Hansen}.  Extending the
classical framework of Brownian motion, as embodied by the Langevin
equation, to relativistic systems is vital to many areas of physics,
which include relativistic fluids/plasmas~\cite{Kovtun,Weldon,Hakim},
effective field theories of dissipative
hydrodynamics~\cite{Grozdanov}, relativistic viscous electron flow in
graphene \cite{Pototsky2012,Marchesoni} and other solids~\cite{Crossno,Fogler}, quark-gluon plasma (QGP) \cite{Hirano} and nuclear
matter~\cite{Boilley}, where thus far only the classical
non-relativistic Langevin-type description is customarily used to
describe fission and fusion-fission processes of hot
nuclei~\cite{Gontchar,Schmidt,Kolomietz2,Kolomietz,Eccles}. Nonetheless,
in all these cases it is clear that relativistic corrections would be
vital.  Furthermore, a Lorentz-covariant extension of the Langevin
equation framework would be the first major step towards building a
consistent relativistic version of nonequilibrium statistical
mechanics~\cite{hang} as well as of fluctuations theorems and thermodynamics in general~\cite{Paraguassu,Pal_2020}. 

One expects that the Langevin model generalizes to the limits of both
fast-moving fluids (where the relativistic motion of constituent
particles becomes important), and for relativistic Brownian
particles. Several relativistic or quasi-relativistic extensions of
the Langevin equation have been proposed, adopted and
used~\cite{Dunkel_review,vojta2020probability,RappVanHees,Quasirel-langev-eq}.
However, rarely if ever any justification beyond that of an ansatz is
provided, and no derivation from first principles is given for the case of
  relativistic environments. While there are many general
  approaches~\cite{hang,dunkel_hanggi} to extending the Langevin
  equation to the weakly- or moderately-relativistic regime, several
  are either postulates of a dissipation
  model~\cite{dunkel_hanggi,RappVanHees,Debbasch}, or generalizations
  of causal fluid dynamics~\cite{causal_hydro} to a case where they
  have limited applicability. Some works provide a justification for
  the case of the relativistic Brownian particle in a non-relativistic
  environment~\cite{vojta2020probability,RappVanHees,Quasirel-langev-eq}. We,
  however, present bottom-up derivations of the relativistic extension
  of the Langevin equation that is valid for relativistic environments
  as well as relativistic particles, and which recovers the
  appropriate results when taking the respective limits.

  Our approach is to extend a derivation of the Langevin equation from
  the Caldeira-Leggett model~\cite{Zwanzig1973,Caldeira_1993}.
  Section~\ref{sec:method} is largely expository; we feel it is
  important to recall the particle-bath non-relativistic treatment in
  detail as the starting point of our derivation. It is useful in
  order to appreciate both the many parallels and the fundamental
  differences that arise with the relativistic treatment.  We then
  (Sec.~\ref{sec:relativistic}) alter the (Galilean) Lagrangian to be
  explicitly Lorentz-covariant. 

  Due to the many difficulties associated with non-linear equations
  that result from the aforementioned Lorentz-covariant Lagrangian,
  the largest portion of the paper is dedicated to obtaining a compact
  approximation to the equations of motion of both the heat bath and
  the tagged particle. This procedure is described in
  Section~\ref{sec:relat-moti-heat}, and a particularly detailed
  breakdown of the necessary steps is given in
  Section~\ref{sec:setup}. Broadly, we obtain a numerical and then a
  quasi-analytical closed-form approximation to the solution of the
  (Euler-Lagrange) equations of motion of the heat bath. We then use
  this solution to construct a coarser, but tractable, approximation
  that is both obviously a relativistic dual of the respective
  intermediate result of the Galilean derivation, and accounts for the
  most important relativistic corrections. By integrating the heat
  bath equations of motion, we obtain the final Lorentz-covariant dual
  of the Langevin equation, which is the central result of the present
  work.

Before we begin, we note that while some authors
(cfr.~\cite{hang,RappVanHees,causal_hydro}) use covariant notation, in
our case, said notation would obscure the intent, and make the
parallels to the non-relativistic derivation harder to see. For the
same reason, we alter the straightforward derivation of the Galilean
Langevin equation to be in the Lagrangian rather than Hamiltonian
formulation.

We shall work with the coordinate time, of one frame referred to as
  the ``lab frame'' in the context of relativistic Brownian
  motion~\cite{hang}. Inasmuch as the classical (generalised) Langevin
  equation derived from the particle-bath Caldeira-Leggett model is
  able to correctly describe Brownian motion in classical statistical
  mechanics, we shall assume that also our results may apply to
  general relativistic Brownian processes, at the level of special
  relativity.  The final important distinction from previous attempts
  at extending classical Brownian motion to relativistic speeds, is
  that we aim to describe the behavior of a single Brownian particle
  (the Langevin picture), rather than of probability distributions
  (the diffusion or Fokker-Planck picture).  Probabilities are rife
  with counter-intuitive phenomena that could be mistaken for causal
  propagation at superluminal speeds~\cite{EPR_paradox}, thus we avoid
  these problems by excluding probabilistic considerations and working within the Langevin framework.

The main result of our paper is a fully-relativistic Langevin equation which describes fully-relativistic motion of both the tagged particle as well as of the environment (the bath oscillators). To our acknowledge this is the first time that such description is achieved, being all previous approaches either valid only for weakly-relativistic systems or for a relativistic particle in a non-relativistic environment.

\section{Galilean Langevin Equation}\label{sec:method}

In this section we recall the derivation of the Langevin equation from
a particle-bath Lagrangian under Galilean relativity. It is important
to recall the derivation with some details as it will serve as the
starting point for the new Lorentz-covariant derivation that we will
present in the next Sections. Also, it is important to recall under
which approximations the Langevin equation can be derived from a
particle-bath model, which we shall elaborate on in this section.

Zwanzig~\cite{Zwanzig1973}, and later Caldeira and
Leggett~\cite{Leggett,Caldeira_1993}, considered a single (tagged)
Brownian particle in linear dynamical coupling with a heat bath of
harmonic oscillators. This choice is justified by three observations:
\begin{enumerate*}[label=(\roman*)] 
\item the coupling is an analytical function and usually sufficiently
  weak to be closely approximated by the linear term in its expansion,
\item the stochastic motion of the environment is represent-able as a
  Fourier transform, and
\item the behavior of the environment is only
  affected by the motion of the particle and no other external
  influence, while
\item  the behavior of the particle can often be
  described by a conservative force, and all dissipation is due to the
  interactions with the heat bath (i.e.~no vector potentials).
\end{enumerate*}
The weak coupling stems from the short-range nature of the
interaction. Averaging the impulses to the time scales of the Brownian
motion effectively reduces the force coming from the environment. This
averaging also sufficiently smooths the function to be amenable to
Fourier analysis.

It is important to note that while our method of using the
  Caldeira-Leggett Lagrangian is remarkably compact, it is neither the
  only, nor the most general derivation of the Langevin equation.  The
  latter title belongs to the Zwanzig-Mori
  formalism~\cite{zwanzig2001nonequilibrium,MoriZwanzig}, wherein only
  determinism of the underlying (microscopic and not necessarily
  classical) equations of motion is required.  In Zwanzig-Mori
  formalism projection operators are used to decouple the motion of
  the entire coupled system into a relevant and irrelevant part, and
  obtain equations of motion for the relevant part. For the
  Caldeira-Leggett model, this step is not necessary, resulting in a
  compact derivation, which proves to be sufficient for our
  purposes.

These assumptions lead to the Caldeira-Leggett
Lagrangian~\cite{Leggett}:
\begin{equation}
  L\!
  =
  \!\frac{m \dot{\mathbf{x}}^2}{2}\!
  -\!V(\mathbf{x})
  +\!\sum_i \!\left[
    \frac{m_{i}\dot{\mathbf{q}}_i^2}{2}\!
    -\!\frac{m_i \omega^2_i}{2}
    {\left(\!\mathbf{q}_i\!-\!\frac{g_i}{\omega^2_i}\mathbf{x}\!\right)\!\!}^2
  \right]\!\!.
  \label{eq:c-l-classical}
\end{equation}
where the \(q_{i}\) are the generalised coordinates of the reservoir
modes, \(\omega_{i}\) --- the natural frequency of the \(i\)-th mode
and \(g_{i}\) the specific coupling between the particle and the
\(i\)-th mode. Finally \(x\) is the position of the tagged
particle. Vectors are indicated with bold, and time derivatives with
over-dots. Also recall, that we have chosen the Lagrangian over the
Hamiltonian formulation~\cite{zwanzig2001nonequilibrium}, in order to
take advantage of the Lagrangian formulation's explicit Lorentz
invariance in later sections.

Ordinarily, the potential energy for the system of harmonic
oscillators depends on \(m_{i}\omega_{i}^{2} q_{i}^{2}\), plus we've
stated that there is a coupling between the modes and the tagged
particle, but one does not see why the tagged particle necessarily
contributes to the potential energy. We impose this contribution,
called the \emph{counter-terms}~\cite{Weiss} so that, the Lagrangian
has manifest space and time translation invariance, as well as both
space and time inversion symmetry. It is the reason why one can assume
the leading order elements of a generalised coupling function to be
linear in the difference in the generalised coordinates and
\(\mathbf{x}\). This ad-hoc induction of symmetry is explored more in
Sec.~\ref{sec:symmetry}.

The Euler-Lagrange equations of motion for the particle and for the
bath oscillators, respectively, are given by:
\begin{subequations}
  \begin{align}
    &m\frac{d^2 \mathbf{x}}{dt^2}
      =
      - \nabla V(\mathbf{x})
      + \sum_i m_i g_i
      \left(\mathbf{q}_i - \frac{g_i}{\omega_i^2}\mathbf{x} \right)\!,
      \label{eq:eqs-of-motion-classical-dp-dt}\\
    &m_i\frac{ d^2 \mathbf{q}_i(t;\mathbf{x})}{dt^2}
      =
      - m_i {\omega_i^2} \mathbf{q}_i(t;\mathbf{x})
      + m_i g_i \mathbf{x},
      \label{eq:heat-bath-momentum}
  \end{align}
\end{subequations}
where, without loss of generality, \( {m_i = 1, \forall i \in N} \). As
we shall see later, there is an equivalent choice, that is more
appropriate to the relativistic derivation, but for the sake of
remaining comparable to the standard derivation, we shall continue
with re-scaling the masses to unity. We should also highlight, that we
treat the dependence of \( \mathbf{q} \) on \( \mathbf{x} \) as a
parametric dependence, rather than functional. By contrast, in the
following section, we shall make this dependence a functional one.

We assume that the trajectory \(\mathbf{x}(t)\) is known, and solve
for the motion of the modes \(\mathbf{q}_i{(t;\mathbf{x})}\), as a
parametric function depending on \(\mathbf{x}(t)\). Using the Laplace
transformation of~\cref{eq:heat-bath-momentum}, one obtains:
\begin{widetext}
  \begin{align}
    \mathbf{q}_i(t) =
    \mathbf{q}_i(0) \cos \omega_i t + \dot{\mathbf{q}}_i(0) \frac{\sin
    \omega_i t}{\omega_i} + g_i \int_0^t ds \, \mathbf{x}(s) \frac{\sin
    \omega_i (t-s)}{\omega_i},
    \label{eq:bath-motion-classical}
  \end{align}
  which when integrated by parts yields
  \begin{align}
    \begin{split}
      \mathbf{q}_i(t) -\frac{ g_i}{\omega_i^2}\mathbf{x}(t) =
      &\left[ \mathbf{q}_i(0) - \frac{ g_i}{\omega_i^2}\mathbf{x}(0) \right]
      \cos \omega_i t + \dot{\mathbf{q}}_i(0) \frac{\sin \omega_i t}{\omega_i} - g_i \int_0^t ds \, \dot{\mathbf{x}}(s)\frac{\cos \omega_i(t-s)}{\omega_i^2}. \label{eq:bath-motion-classical-int}
    \end{split}
  \end{align}
  The term,
  \begin{align}
    \begin{split}
      \mathbf{F}_p(t)
      =
      \sum_i g_i m_i
      \Bigg\{
      \left[
        \mathbf{q}_i(0) - \frac{g_i}{\omega_i^2}\mathbf{x}(0)
      \right]
      \cos \omega_i t + \frac{\dot{\mathbf{q}}_i(0)}{\omega_i} \,
      \sin \omega_i t
      \Bigg\},
      \label{eq:noise-classical}
    \end{split}
  \end{align}
\end{widetext}
is known as the \emph{stochastic force}~\cite{zwanzig2001nonequilibrium}.

Substituting \cref{eq:bath-motion-classical-int} into
\cref{eq:eqs-of-motion-classical-dp-dt} we obtain the (Galilean)
\emph{generalised Langevin equation}~\cite{Zwanzig1973,zwanzig2001nonequilibrium}
\begin{align}
  \begin{split}
    m\frac{\,d\mathbf{x}(t)}{dt}\!= \mathbf{F}_p(t) - \nabla
    V\left(\mathbf{x}\right) -
    \int_0^t\!\!\!\dot{{\mathbf{x}}}(t\!-\!s)K(s)
    ds,\label{eq:langevin-classical}
  \end{split}
\end{align}
where
\begin{equation}
  K(t) = \sum_i \frac{ g_i^2}{\omega_i^2} \cos \omega_{i}t,\label{eq:memory-classical}
\end{equation}
is known as the \emph{friction kernel} or the \emph{memory
  function}~\cite{zwanzig2001nonequilibrium}. It represents the
ability of the medium to respond to temporally distributed
perturbations, such as those encoded in \(\mathbf{\dot{x}}\).
\begin{change}
\end{change}

The equation~\eqref{eq:langevin-classical} is called
\emph{generalised}, because we have not yet imposed any restrictions
on the behavior of the contributing terms: the stochastic force and
the friction kernel. The following is an example of obtaining a
special case known as the \emph{Markovian Langevin equation} from
\cref{eq:langevin-classical}. We start by noticing
that~\cref{eq:memory-classical} is a Fourier series decomposition of
\(K(t)\) in terms of \(\omega_i\), as is \cref{eq:noise-classical} a
decomposition of \(F_{p}\). This series decomposition, can be replaced
with a Fourier transform, for a continuous vibration spectrum, and a
given density of states \(\rho(\omega)\). The specific choice of
\(\rho(\omega)\) dictates the behavior of \(K(t)\) and \(F_{p}\), via:
\begin{equation}
  K(t)
  =
  \int_0^\infty \rho(\omega) \frac{g{(\omega)}^2}{\omega^{2}}\cos \omega t \,
  d\omega.
  \label{eq:memory-classical-ft}
\end{equation}
For the Markovian equation, \({\rho(\omega) \propto \omega^2}\), as
per Debye's law and/or the Rayleigh-Jeans law for bosonic baths; and
\({g(\omega)=\text{Const}}\). Thus \(K(t) \propto \delta(t)\) as shown
in~\cite{zwanzig2001nonequilibrium}. The friction term
in~\cref{eq:langevin-classical} becomes the classical viscous
force~\( \eta \dot{x}\) and \cref{eq:langevin-classical} reduces to
the special case of the Markovian Langevin
equation~\cite{zwanzig2001nonequilibrium}.

This derivation not only obtains the Langevin equation
\cref{eq:langevin-classical} from general assumptions: weak coupling,
space-translation invariance; but also provides a clear physical model
for the origins of physical effects such as viscosity and the
stochastic excitation. The most natural model for the environment ---
the harmonic heat bath, is also the same model used for describing
quantum decoherence in the theory of open quantum
systems~\cite[p.~656]{Breuer}.

Importantly, this derivation demonstrates the classical
  deterministic worldview of statistical mechanics.  Even in its final
  form the Galilean Langevin equation would be fully deterministic
  provided knowledge about initial conditions for all the degrees of
  freedom is at hand~\cite{zwanzig2001nonequilibrium}.  The memory
  function and the stochastic force can be fully predicted both
  forwards and backwards in time, provided the initial conditions at
  an arbitrary \(t=0\) are known. The reason why the memory function
  and the stochastic force \emph{appear} to break time-inversion
  symmetry, is because
  \begin{enumerate*}[label=(\roman*)] 
  \item The number of Fourier modes (i.e. bath oscillators) is very large (in fact, merely
    recording \(O(10^9)\) initial conditions already requires at least
    \texttt{16Gib} of RAM) and their initial conditions cannot be specified deterministically, hence they have to drawn from \emph{statistical} distributions such as Boltzmann or J\"{u}ttner for the non-relativistic and relativistic limits respectively.\label{it:many}
  \item The system is possibly chaotic due to the intrinsic non-linearity of the relativistic oscillators: minor perturbations to the initial
    conditions can greatly impact long-term
    behavior.\label{it:chaotic}
  \end{enumerate*}
  As a result, while \cref{eq:langevin-classical} is technically
  deterministic, it is \emph{effectively} stochastic \cite{Zwanzig1973,zwanzig2001nonequilibrium}.  

  Finally, one must note the fluctuation-dissipation
  relations~\cite[see p.~23]{zwanzig2001nonequilibrium}:
  \begin{align}
    \left\langle \mathbf{F}_{p}(t)\right\rangle = &0,\\
    \left\langle \mathbf{F}_{p} (t) \mathbf{F}_{p} (t') \right\rangle = &k_{B}T K (t - t'),
  \end{align}
  where \(\langle \ldots \rangle\) denotes ensemble or time averaging
  (which are equivalent for ergodic fluids).

The Lagrangian~\eqref{eq:c-l-classical} is manifestly Galilean
covariant and not Lorentz covariant. What this means is that the
\cref{eq:langevin-classical} is not accurate in cases where either the
environment and/or the tagged particle, are highly
relativistic: e.g.~stellar cores, the early universe, as well as the
plasma in collider experiments.

\section{Relativistic Caldeira-Leggett Lagrangian}\label{sec:relativistic}

  In this section we update the Caldeira-Leggett model and the
  derivation of the generalised Langevin equation to work in special
  relativity. While there are relatively many descriptions of a
  relativistic Brownian particle coupled to a non-relativistic heat
  bath~\cite{hang,dunkel_hanggi}, there are far fewer attempts at tackling
  the relativistic behaviour of the heat bath. Recall that in the
  Galilean case, the main difficulty lies in obtaining the equations
  of motion for the heat bath modes. This task is non-trivial, as
  there are many relativistic Lagrangians that reduce to
  \cref{eq:c-l-classical} in the relevant limit.

  Moreover, in the previous consideration we have encapsulated all
  external forces in
  \({V_\text{ext}(\mathbf{x}) = \phi(\mathbf{x})}\), which has
  causality implications under special relativity.  By analogy with
  electromagnetism, we instead consider
  \begin{align}
    V(\mathbf{x}, \mathbf{\dot{x}}, t)
    =
    \phi (x, t)
    - \frac{\mathbf{\dot{x}} \cdot \mathbf{A} }{c}
  \end{align}
  adding an extra degree of freedom in \(\mathbf{A}\) and re-absorbing
  the ``charge'' equivalent into the definitions of \(\mathbf{A}\) and
  \(\phi\). Since electromagnetism is Lorentz covariant, we get
  familiar equations of motion and sidestep problems with
  action-at-a-distance.

  In the original Galilean derivation by Zwanzig~\cite{Zwanzig1973},
  the choice of harmonic equations of motion is a reflection of the
  efficacy of Fourier methods. It might be tempting to carry over the
  harmonic trajectory, rather than modifying the interaction. However
  this is mathematically equivalent to non-relativistic heat baths
  already considered in~\cite{hang,dunkel_hanggi}. Thus, instead, the
  heat bath dynamics are relativistic equivalents of harmonic
  oscillations, rather than harmonic oscillations themselves.  There
  are many models which reduce to simple harmonic oscillators, many of
  which are not solvable~\cite{MacColl,Babusci,Schoberl}. We chose 
  \begin{align}
    \begin{split}
      ds^2\left( q_{i}, x\right)
      =
      &c^{2} {\left( \!t_{i}\! -\!\frac{g_{i}}{\omega_{i}^{2}}t\!\right)}^{2}\!\!
      - \!{\left(\mathbf{q}_{i} - \!\frac{ g_{i}}{\omega_{i}^{2}}
          \mathbf{x} \right)}^{2}\!\!, 
    \end{split}
  \end{align}
  to parameterise the relativistic equivalent of the harmonic
  potential, for the following reasons:
  \begin{enumerate*}[label=(\roman*)] 
  \item it is obviously a norm of a 4-vector,
  \item it obviously corresponds to the conventional coupled harmonic
    potential in the limit \(c \rightarrow \infty\)
  \item the constituent ``events'' can straightforwardly be
    interpreted as an interaction mediated by a (massless) particle,
  \item requiring that the ``events'' be null-separated is the
    relativistic (local) equivalent of Galilean (global) simultaneity, 
  \item when the interaction ``events'' are null-separated, many of
    the mathematical complications vanish. 
  \end{enumerate*}

\begin{figure} \centering
  \includegraphics[width=\columnwidth]{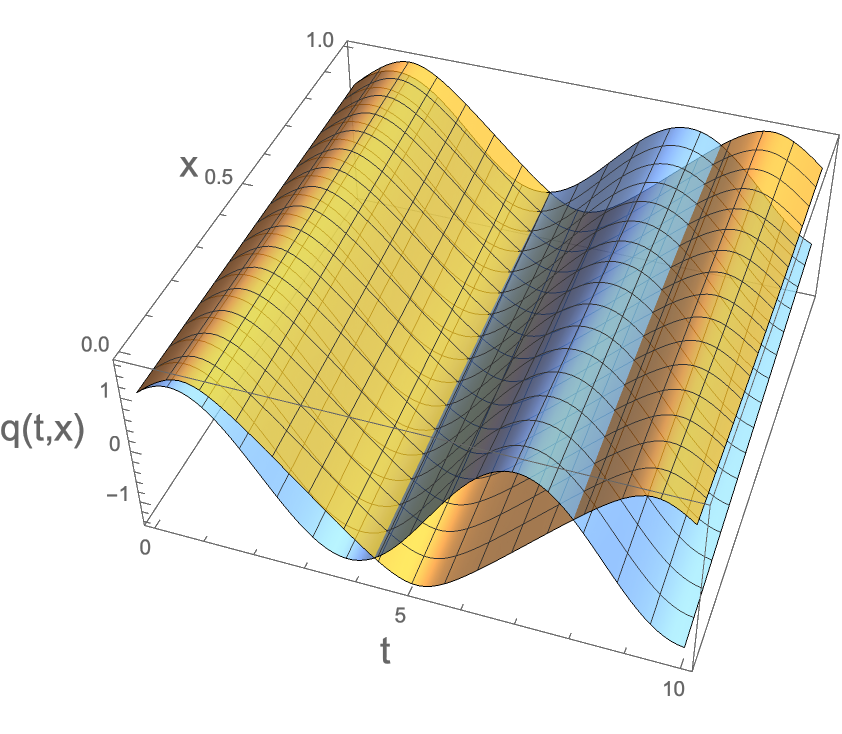}
  \caption{Comparison of the Galilean analytical (blue) and the
    relativistic numerical (orange) solutions of heat bath equations
    of motion. Note that the mismatch stems from a difference of
    frequency and not of phase. This leads to a renormalised frequency
    in
    \cref{eq:bath-motion-relat}. }\label{fig:galilean-relativistic-comp}
\end{figure}

  Bringing all of these considerations together we arrive at
\begin{align}
  \label{eq:starting-lagrangian-rel}
  \begin{split}
    L\!
    =
    &\gamma^{-1}(\dot{\mathbf{x}}) mc^{2}
    - V(\mathbf{x}, \mathbf{\dot{x}}, t)
    + \sum_i m_i\gamma^{-1}(\dot{\mathbf{q}}_{i})c^{2}\\
    &+\! \sum_{i}\!\frac{ m_{i}\omega^{2}_{i}}{2\gamma( \mathbf{\dot{x}})}
    \left[
      c^2{\left(t_i - \frac{g_i}{\omega_{i}^{2}}t \right)}^2\!
      - {\left(\mathbf{q}_i - \frac{g_i}{\omega^{2}_{i}} \mathbf{x} \right)}^{2}
    \right]\!\!,
  \end{split}
\end{align}
where
\begin{equation}
  \gamma(\mathbf{v}) = \frac{1}{\sqrt{1-
      \frac{|\mathbf{v}|^{2}}{c^{2}}}},\label{eq:def-gamma}
\end{equation}
and overdots signal differentiation with respect to corrdinate
time. We must note the counter-intuitive presence of \(\gamma^{-1}\)
rather than \(\gamma\) in the kinetic terms. This term in the
Lagrangian has been discussed at length in standard
literature~\cite[p.~323]{goldstein1980classical}. It pertains to the
difference between proper time and coordinate time, and the fact that
the Euler-Lagrange equations are formulated using proper, rather than
coordinate time (see~\citet[p.~320]{goldstein1980classical}).  The
coupling term as discussed previously is formulated in terms of the Lorentz-invariant 4-interval
  between discrete interaction events (\cref{eq:starting-lagrangian-rel}), following the same logic as the
introduction of counter-terms in the original non-relativistic
Caldeira-Leggett Lagrangian~\cite{Caldeira_1993,Weiss}. 

The equations of motion associated to the
Lagrangian~\eqref{eq:starting-lagrangian-rel} are
\begin{subequations}
  \begin{align}
    &\frac{d}{d t} [\gamma(\dot{\mathbf{x}}) m
      \dot{\mathbf{x}}]
      =
      \sum_{i} \frac{m_{i}}{\gamma(\mathbf{\dot{x}})}  g_{i}\left(
      \!\mathbf{q}_{i}\!-\!\frac{g_{i}}{\omega_{i}^{2}}\mathbf{x}\!
      \right)\!
      -\mathbf{F}_{\text{ext}}
      \label{eq:eom-rel-x}, \\
    &\frac{d}{d t}
      [\gamma(\dot{\mathbf{q}}_{i}) m_{i} \dot{\mathbf{q}}]
      =
      - \frac{m_{i}}{\gamma(\mathbf{\dot{x}})} \omega_{i}^{2}\mathbf{q}_{i}
      + \frac{m_{i}}{\gamma(\mathbf{\dot{x}})} g_{i}\mathbf{x}
      \label{eq:eom-rel-q},
  \end{align}
\end{subequations}
where the functions are evaluated at coordinate time \(t\) and
\({\mathbf{F}_{\text{ext}} = \nabla \phi(t,\mathbf{x}) + \nabla
  (\frac{\mathbf{A} \cdot \mathbf{\dot{x}}}{c}) -
  \frac{\dot{\mathbf{A}}}{c}}\). In the equations of motion, this is
the only term due to differentiation of the potential with respect to
the first derivative that survives. The terms resulting from
differentiating \(\gamma(\dot{\mathbf{x}})\) contribute a full
interval, which is null in our model. The factors of \(\gamma^{-1}\)
signal that the Lorentz contraction cannot and does not contribute to
the force.

\section{Heat bath mode trajectories}\label{sec:relat-moti-heat}
\subsection{Problem setting}\label{sec:setup}
We proceed in complete analogy with the Galilean case, assuming that
\(\mathbf{x}(t)\) is a known trajectory, and reverse-engineer the
equation satisfied by the heat bath. By expanding the left hand side
of \cref{eq:eom-rel-q} and using the product rule, we obtain
\begin{equation}
  \gamma(\dot{\mathbf{q}}) m_i \mathbf{\ddot{q}}
  - \gamma^3(\dot{\mathbf{q}})m_i \,(\mathbf{\ddot{q}}
  \cdot \dot{\mathbf{q}}) \dot{\mathbf{q}}
  =
  \frac{m_{i}}{\gamma( \mathbf{\dot{x}} )} g_{i}\mathbf{x}
  - \frac{m_{i}}{\gamma(\mathbf{\dot{x}})} \omega_{i}^{2}\mathbf{q}_{i}.
  \label{eq:correct-vector}
\end{equation}

  To proceed, one projects \( \ddot{\mathbf{q}} \) along the direction
\(\dot{\mathbf{q}}\), decoupling the trajectories into longitudinal
and transverse components.

By combining similar projections we obtain%
\begin{subequations}
  \begin{equation}
    \gamma^{3} (\dot{\mathbf{q}}_i) m_i {\ddot{\mathbf{q}}_{i}}{}_{||}(t, \mathbf{x})\!
    =\frac{m_i}{\gamma(\mathbf{\dot{x}})} g_i {\mathbf{x}}{}_{||}\!
    - \frac{m_i}{\gamma(\mathbf{\dot{x}})} \omega_i^2 {\mathbf{q}_i{}_{||} },\label{eq:N2-relat}
  \end{equation}
  for the longitudinal, 
and
  \begin{equation}
    \gamma(\dot{\mathbf{q}}) m_i {\ddot{\mathbf{q}}_i}{}_{\bot}(t, \mathbf{x})
    =
    \frac{m_i}{\gamma(\mathbf{\dot{x}})} g_i \mathbf{x}{}_{\bot}
    - \frac{m_i}{\gamma(\mathbf{\dot{x}})} \omega^2_i\mathbf{q}_i{}_{\bot},
    \label{eq:N2-relat-2}
  \end{equation}
\end{subequations}
for the transverse component. Here, notice that we have used
\(\mathbf{q}(t, \mathbf{x})\),
  while
in reality \(\mathbf{q}\) is only a
function of time, and the \(\mathbf{x}\)
  dependence comes indirectly from assuming knowledge of the
  trajectory \(\mathbf{x}(t)\). Viewing this implicit dependence as an
  explicit depdendence on a free variable \(\mathbf{x}\) is what
  allows us to solve \cref{eq:N2-relat-2,eq:N2-relat}.

  Our focus for this section shall be developing a method for solving
  the two equations. At present, mathematics does not support an exact
  closed-form solution in terms of standard
  functions~\cite{Babusci,Schoberl,RHO} to either of the known
  relativistic harmonic oscillators.  Since \cref{eq:N2-relat-2} and
  \cref{eq:N2-relat} break the isotropy of the problem, one would
  expect that the components of the solution need to be tracked
  separately. However, looking at the McLaurin series expansion 
  \begin{equation}
    {\ddot{ \mathbf{q}}_i}{}_{||}\!=\!  \frac{1}{\gamma(\mathbf{\dot{x}})}\left( g_i
      \mathbf{x}{}_{||}\!-\omega_i^2 \mathbf{q}_i{}_{||}\! \right)
    \left[1\!+\!\frac{3\left| \dot{\mathbf{q}}_i{}_{||}\right| ^2}{2c^2}\!+\!O
      \left( \frac{\left| \dot{\mathbf{q}}_i{}_{||}\right| ^{4}}{c^{4}} \right)
    \right]\!\!,\label{eq:N2-relat-approx}
  \end{equation}
  of \cref{eq:N2-relat} in the limit
  \({\mathbf{\dot{q}} \rightarrow 0}\) and comparing it to the
  expansion of \cref{eq:N2-relat-2}, we can see that the only
  difference is the numerical factor of powers of
  \({\mathbf{\dot{q}}}/{c}\). This suggests that a functional form
  approximating the solution to \cref{eq:N2-relat-2} can also
  approximate \cref{eq:N2-relat}, albeit with different numerical
  parameters. We shall be working under this assumption, and justify
  it in Appendix~\ref{sec:bayes-just-prop}.

  Our plan for solving the relativistic heat bath's
  equations of motion is as follows. Firstly, we shall produce a
  numerical solution to a simplified \(1+1\) dimensional problem, using
  the longitudinal component --- \cref{eq:N2-relat},  as a
  base (Sec.~\ref{sec:numerics}). We shall then use this numerical solution to construct a more compact 
  closed-form solution that retains cardinal properties of the numerical solution,
  while remaining sufficiently similar to \cref{eq:bath-motion-classical-int} (Sec.~\ref{sec:numeric-to-analytic}). As it happens, 
  \cref{eq:N2-relat} can be solved in terms of elliptical functions, and while the solution is too cumbersome to even quote (much less manipulate), we shall use some of its properties to validate our previous step (Sections~\ref{sec:numeric-to-analytic} and \ref{sec:integr-const-init}). In Sec.~\ref{sec:integr-const-init} we find (and in Appendix~\ref{sec:bayes-just-prop} verify), that the closed form approximation requires parameter fitting in order to be a good approximation to the solution of \cref{eq:N2-relat}. We carry out said fitting in Sec.~\ref{sec:matching}, and explain the necessary modifications to apply the same process to \cref{eq:N2-relat-2} in Sec.~\ref{sec:transverse}.

\begin{figure}[b]
  \centering%
  \includegraphics[width=\columnwidth]{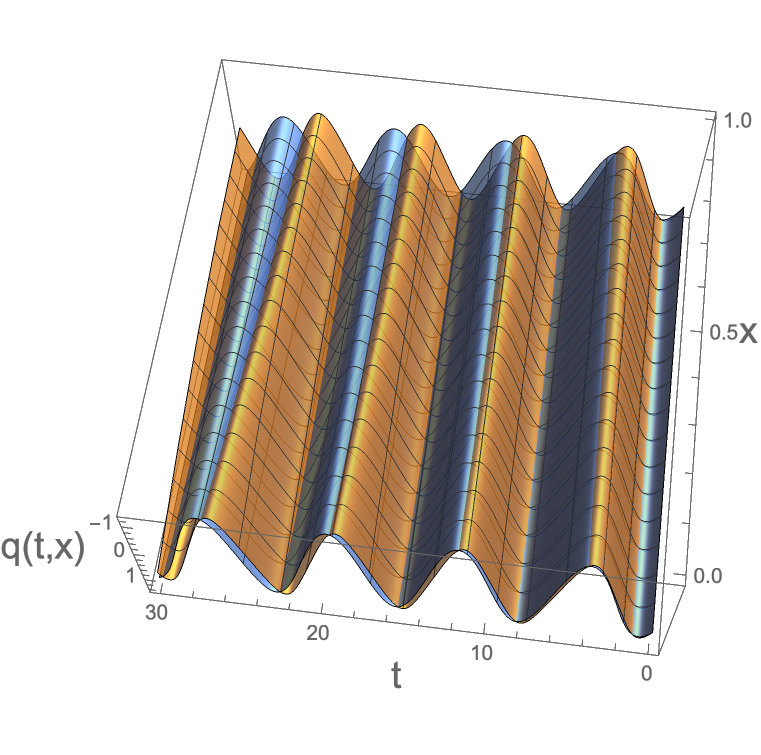}
  \caption{Comparison of frequency-matched (by setting
    \(\bar{\omega} = \text{Const.}\), \(\xi = \text{Const.}\)) version
    of~\cref{eq:bath-motion-relat} (blue) to the numerical solution
    of~\cref{eq:N2-relat} (orange). The divergence at higher values of
    \(x\), illustrates the necessity of introducing
    \(\xi(\mathbf{x}, t)\).  }\label{fig:x-dependence}
\end{figure}

\begin{figure}[b]%
  \centering%
  \includegraphics[width=\columnwidth]{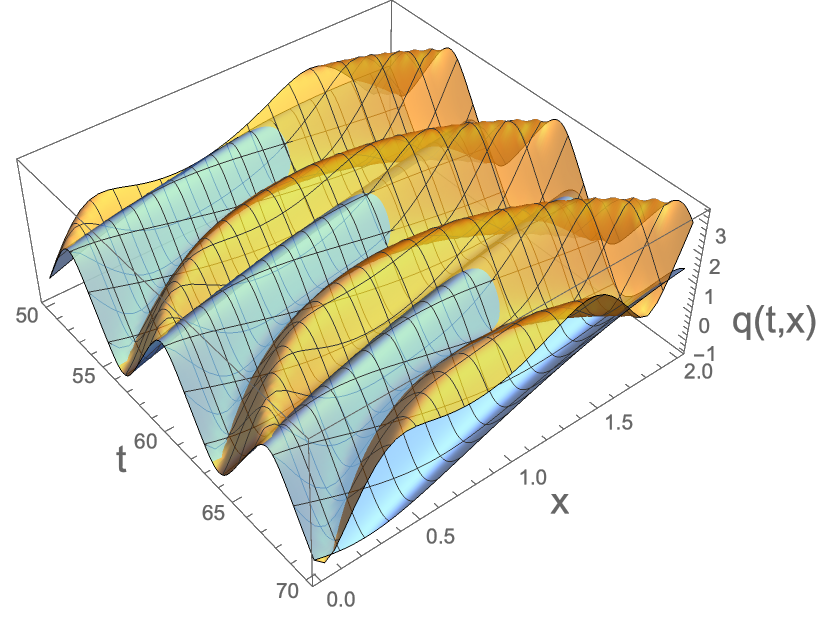}
  \caption{A 3D plot showing the increasing deviation of the heat bath
    trajectory (orange) from a plane wave (blue) with the same base
    frequency, i.e.~fitted \(\xi = \text{Const.}\) and
    \(\bar{\omega}\).}\label{fig:xi-3d}
\end{figure}

\subsection{The numerical solution\label{sec:numerics}}
By comparing the numerical solutions of~\cref{eq:N2-relat} and
\cref{eq:bath-motion-classical}
in~\cref{fig:galilean-relativistic-comp}, we can see that relativistic
corrections manifest as
\begin{enumerate*}[label=(\roman*)] 
\item a shift in eigenfrequency of plane wavelets (see
  \cref{fig:galilean-relativistic-comp}) of the bath oscillators.
\item a slight dependence of the wavelet phase of the bath oscillators
  on the position of the tagged particle \(\mathbf{x}\) (see
  \cref{fig:xi-contour,fig:xi-3d,fig:x-dependence}).
\item smaller sub-oscillations along contours of constant phase,
  (\cref{fig:xi-3d}). This effect is weak, and ignoring it only
  slightly reduces the \(\chi^{2}\) of the numerical fit, but greatly
  reduces the complexity of the following manipulations.
\end{enumerate*}
So in order to build a Lorentz-covariant analogue of
\cref{eq:bath-motion-classical}, we will need to incorporate these
features.

We do this by introducing an adjusted or renormalised frequency
\(\bar{\omega}\), and a relativistic-correction phase
\(\xi(\mathbf{x}(t))\) into the sinusoidal forms of the bath dynamics.
These two corrections play a crucial role in building the solution to
the dynamical problem, and they have a clear physical origin as
summarised below.

The renormalization of the oscillator frequencies \(\omega_i\) is
entirely expected since the dynamics of the relativistic oscillator is
nonlinear (see \cref{eq:eom-rel-q} versus
\cref{eq:bath-motion-classical}). As we know from classical mechanics,
the eigenfrequency of an anharmonic oscillator is given by the
eigenfrequeny of the corresponding harmonic oscillator plus a
correction which depends on the anharmonic coefficients (as well as
the square of the amplitude), see e.g.~\cite[pp.87-88]{Landau}.

Clearly, in our case the harmonic eigenfrequency is the one of the
non-relativistic oscillator (\(\gamma(\dot{q})=1\)),
i.e. \(\omega_{i}\), whereas the renormalised eigenfrequency
\(\bar{\omega}_{i}= \omega_{i} + \delta\omega_{i}(\gamma(\dot{q}))\)
is given by \(\omega_{i}\) plus a correction \(\delta\omega_{i}\) that
depends on the degree of anharmonicity (nonlinearity) of the
system~\cite{Landau}, i.e.~on \(\gamma(\dot{q})\) in our case. We indeed found from numerical simulations for the initial conditions that we investigated that
\(\bar{\omega} \sim \gamma^{-3/4}(\dot{q})\).

The relativistic-correction phase \(\xi\), instead, is required in
order to bring the oscillatory/sinusoidal terms in the solution into a
plane wave form. In turn, the so-obtained plane wave form is
Lorentz-covariant thanks to Lorentz-invariance of the wave equation
under the Lorentz transformations~\cite[p.~383]{Jackson}. We shall
provide a detailed clarification and analysis of these terms in the
next section.

We want our solution to be a manifestly Lorentz-covariant version
of~\cref{eq:bath-motion-classical}. Motivated by the above physical
considerations, and upon incorporating the mentioned corrections, we
obtain:
\begin{align}
  \begin{split}
    \mathbf{q}_i(t)
    =
    &\mathbf{q}_i(0) \cos \bar{\omega_i}
    \left(t - \frac{\bar{\xi}_{i}(t, \mathbf{x}(t))}{c}\right)\\
    &+ {
      \gamma(\dot{\mathbf{q}}_{i} (0))\mathbf{\dot{q}}_i(0)
      \frac{\sin \bar{\omega}_i \left( t - \frac{\bar{\xi}_{i}(t, \mathbf{x}(t))}{c} \right)}
      {\bar{\omega}_i}
    }\\
    &+ {
      g_i \!\! \int_0^t \!\!\!\! \gamma(\mathbf{\dot{x}}(s)) \mathbf{x}(s)
      \frac{\sin \bar{\omega}_i
        \left(t - \frac{\bar{\xi}_{i}(t, \mathbf{x}(s))}{c} - s \right)}
      {\bar{\omega}_i}\,
    }ds,
    \label{eq:bath-motion-relat}
  \end{split}
\end{align}
where \(\bar{\omega}_{i}\) is the renormalised frequency that we
introduced above,~\footnote{which is a necessary consequence of the
  nonlinear, anharmonic character of the relativistic
  oscillator~\cite{Landau}}. This analytical form provides a
reasonably accurate, Lorentz-covariant match of the numerical
solutions to \cref{eq:N2-relat}, thus eliminating the mismatch shown
in~\cref{fig:galilean-relativistic-comp}.

Firstly, unlike a na\"{\i}ve extension of the
Galilean~\cref{eq:bath-motion-classical},
in~\cref{eq:bath-motion-relat}, we have locality and causality
encapsulated in the phase \(\bar{\xi} (t,\mathbf{x}(t))\), which is
an integration constant. To understand the origin and physical meaning
of this parameter, one must delve deeper into the solution's
structure.

Secondly, we have two pairs of functions: \(\hat{\xi}(\mathbf{x})\)
and \(\bar{\xi}(t, \mathbf{x}(t))\), which we shorten to
\(\bar{\xi}(t)\); and \(\hat{\omega}\) and \(\bar{\omega}\). The first
pair represents the phase for two distinct situations, one is
dependent only on \(\mathbf{x}\), while the other also explicitly
depends on \(t\). Similarly, the second pair are the frequencies for
the same situations. The two formulations are equivalent as shown in
Appendix~\ref{sec:tdfrequency}. We've added the overbar to \(\bar{\xi}\) for consistency.

\subsection{Computationally-informed structure of the solution}\label{sec:numeric-to-analytic}

In this section, we  provide a justification of the
aforementioned correction coefficients, i.e.~the renormalised
frequency \(\bar{\omega}\) and phase \(\bar{\xi}\), in reference
to a quasi-analytical solution of~\cref{eq:N2-relat} obtained in terms of elliptical functions in $1+1$
dimensions.  For compactness the indices \(i\) are dropped and
reinserted only in \cref{eq:bath-motion-relat-int}.  The
computationally-obtained quasi-analytical solution is too long to be
reported here (it can be found in the accompanying Mathematica
notebooks~\cite{mathematica}). Instead we use it to study the generic
features of the solution in order to understand and justify the
compact, analytical approximation, as well as 
explain the origins and methods of obtaining the renormalised
frequency \(\bar{\omega}\) and a time dependent phase \(\xi\) from initial conditions.

We consider \(\mathbf{x} = x\) as an independent variable of the
scalar function \(q(t, x)\), and attempt to solve~\cref{eq:N2-relat},
as if it were an ordinary differential equation in \(t\), allowing
\(x\) to vary.  This allows for one extra degree of freedom, which
permits approximating the solution in terms of elliptical
functions. This solution, though cumbersome, contains two integration
constants, whose roles are closely related to the roles of
\(\bar{\omega}\) and \(\bar{\xi}\) in
\cref{eq:bath-motion-relat}.
Firstly, the generic form of
the Mathematica-generated quasi-analytical solution reads as
\begin{equation}
  \mathbf{q}(t, \mathbf{x})
  =
  F\left(t - C_{2}(\mathbf{x}); C_1 (\mathbf{x})\right),
  \label{eq:exact-top-level}
\end{equation}
  where \(F\) is an expression whose exact structure is not
  important~\cite{mathematica}, save for being quasi-periodic.  It is
  only approximately periodic in \(t\) i.e.~no longer independent of
  \(\mathbf{x}\), as in \cref{eq:bath-motion-classical}, (see
  \cref{fig:xi-contour}), due to the presence of \(C_2(\mathbf{x})\)
  in the solution (and \(C_{1}(\mathbf{x})\) to a lesser extent). By
  comparing \cref{eq:bath-motion-relat} and \cref{eq:exact-top-level}
  and matching the arguments at the local maxima of both functions,
  \(\hat{\xi}(\mathbf{x})/c = C_2(\mathbf{x})\) must be the
  integration constant responsible for the curvature of the wavefronts
  in the solution as \(\mathbf{x}\) increases (see~\cref{fig:xi-3d}).

  To understand the relationship between the other integration
  constants, one must dig more deeply into the structure of \(F\).
  The form of \(F\) is the inverse of a combination of elliptical
  functions of a compound argument, that involves
\(C_1(\mathbf{x}), t - C_2(\mathbf{x}), \omega\) and \(g\). Neither
the expressions that contain \(C_1(\mathbf{x})\) nor \(\omega\) can be
factored out. Thus when formulating \cref{eq:bath-motion-relat}, we
reflect this by introducing
\(\hat{\omega} = \hat{\omega}(g\mathbf{x}, \omega, t, \ldots)\), which
encapsulates the fact that the solution is anharmonic, and that the
frequency is shifted by a quantity that depends on
\(C_{2}(\mathbf{x})\).
Thus both the phase and the
  frequency are position-dependent. This introduces an unnecessary
  complication which we can avoid by defining our integration
  constants slightly differently by exploiting a degree of freedom
  (see Appendix~\ref{sec:integr-const-init}). For our purposes it is
  easier to work with a time independent \(\bar{\omega}\) with all of
  its time (and position) dependence added to \(\bar{\xi}\), rather
  than with \(\hat{\omega}\) and \(\hat{\xi}\).

Both \(\bar{\xi}\) and \(\bar{\omega}\) play the role of integration
constants. In principle, if one had a set of initial conditions to fix
both \(C_1(\mathbf{x})\) and \(C_2(\mathbf{x})\), one could solve the
equation 
\begin{equation} 
\mathbf{q}(t,\mathbf{x}) = \mathbf{F}(t - C_2(\mathbf{x}); C_1(\mathbf{x})),
\end{equation} 
where \(\mathbf{q}(t,\mathbf{x})\) is defined in
\cref{eq:bath-motion-relat}, to obtain both \(\bar{\xi}\) and
\(\bar{\omega}\) in closed form as functions of \(C_{1}\) and
\(C_{2}\).
Notice, that there are two degrees of freedom
  corresponding to \(C_1(\mathbf{x})\) and \(C_2(\mathbf{x})\) that
  transform into two degrees of freedom in \(\bar{\xi}\) and
  \(\bar{\omega}\), that are linked with only one equation. This is
  what allows us the freedom to choose to work with \(\bar{\xi}\) and
  \(\bar{\omega}\), rather than \(\hat{\xi}\) and \(\hat{\omega}\),
  which are more directly related to integration constants of
  \cref{eq:exact-top-level}.

We also remind the reader that \(\mathbf{x}\) enters expressions that
straightforwardly generalize to (1+3) dimensional dynamics, while
expressions containing \(x\) are specific to (1+1) dimensional
considerations. Case in point, the concept of wavefronts is a
convenient representation of the Fourier decomposition of the motions
of \(\mathbf{q}\) and \(\mathbf{x}\) that only applies to the case of
scalar \(x\) and \(q\).

\begin{change}
\end{change}

\subsection{Integration constants and initial
  conditions}\label{sec:integr-const-init}

At this point it might appear that the phase-type term
\(\bar{\xi}(\mathbf{x})\) can be chosen arbitrarily, but that is not
the case. We are bound by causality, which links \(C_{1}\) and
\(C_{2}\), and gives us a clear understanding of the functional form
of \(\mathbf{q}_i (\mathbf{x}, t)\). By looking briefly at the
functional form of \(F(\lambda) \) returned by
Mathematica~\cite{mathematica}, one shall notice a deluge of terms,
many of which fall into either
\begin{subequations}
  \begin{equation}
    \sqrt{2 \omega^4 C_1(\mathbf{x})+g^2 x^2+2\omega^2},\label{eq:A-term}
  \end{equation}
  or
  \begin{equation}
    \sqrt{2 \omega^4 C_1(\mathbf{x})+g^2 x^2-2 \omega^2},\label{eq:Astar-term}
  \end{equation}
\end{subequations}
where we used the shorthand \(x = \left| \mathbf{x} \right| \). While
it needs to be shown conclusively and rigorously, it is rather evident
that causality constrains all such terms to be real.

The solution also contains terms that mix the argument, (thus
\(C_2(\mathbf{x})\)) with \(C_1(\mathbf{x})\) under a square root, so
we can conclude that the integration constants are not independent of
each other.  An imprint of causality should be present on
\(C_{2}(\mathbf{x})\) as well as \(C_1(\mathbf{x})\).

Finally, we must reconcile the integration constants, \(C_{1}\) and
\(C_{2}\), with the constant boundary conditions that we have assumed
for the Galilean case.  The constant boundary conditions could violate
causality, thus proving problematic. So instead of assuming that the
boundary conditions are constant across the entire domain of the
solution, we instead assume that the boundary conditions are constant
in the domain of events which are separated from the co-ordinate
origin in a time-like fashion, i.e.~within the light-cone of the
origin, and zero otherwise.

As it turns out (see Appendix~\ref{sec:gx}), the relevant separation
scales at which this could occur, are far in excess of the distances
at which other assumptions we've made would break down. So we are
justified in constraining the domain of \( x \in (0,1) \), (\( g=2 \)
for convenience) and in choosing arbitrary initial conditions. For
example,
\begin{subequations}\label{eq:bc}
  \begin{align}
    \mathbf{q}(0, \mathbf{x}) = &1 \label{eq:bc-q}\\
    \dot{\mathbf{q}}(0, \mathbf{x}) = &0.85, \label{eq:bc-qdot}
  \end{align}
\end{subequations}
within the light-cone of the current event
\( \mathbf{q}(0, \mathbf{0}) \). Here we chose both numbers and
constant initial conditions out of convenience, and for illustration
purposes.  \( \mathbf{q}(0, \mathbf{x}) =1 \) is an arbitrary choice,
while in the units of \( c=1 \),
\( \dot{\mathbf{q}}(0, \mathbf{x}) = 0.85\) is both sufficiently large
to show relativistic effects, but also the largest number that doesn't
result in numerical errors.

Now suppose that we have imposed the relevant initial conditions and
that the conditions are valid for our particle-bath system. We can
thus obtain \(C_{1}(x)\) and \(C_{2}(x)\) from the closed-form
approximation to the numerical solution detailed above. Moreover, we
can determine what the \(\bar{\xi}(\mathbf{x}, t)\) is for the
relevant \(\bar{\omega} = \text{Const.}\) via algebraic manipulations
of elliptic and trigonometric functions, at least in
principle. However, it is more relevant to attempt a different
approach, that we have used to generate the relevant plots
illustrating the match between the numerical solution and
\cref{eq:bath-motion-relat}.

\subsection{Matching \cref{eq:bath-motion-relat} to the numerical  solution}\label{sec:matching}
By considering \cref{eq:bath-motion-relat} subject to initial
conditions \cref{eq:bc} along the line \(\mathbf{x} = \mathbf{0}\), we
see that \cref{eq:bath-motion-relat} is a generalised sinusoidal
function, with the caveats for \(\bar{\xi}\) and \(\bar{\omega}\)
discussed above. The true solution to \cref{eq:N2-relat} (and to
\cref{eq:N2-relat-2}), however, is not a perfect sinusoid, but can be
approximated arbitrarily well by a suitable choice of \(\bar{\xi}\)
and \(\bar{\omega}\), which one can show by substituting
\cref{eq:bath-motion-relat} into \cref{eq:N2-relat} (and
\cref{eq:N2-relat-2}). Of course, we are interested in the opposite, finding
\(\bar{\xi}\) and \(\bar{\omega}\) that minimize the mismatch.

Our first intuition is that the solution is fully periodic, so the
frequency at \(\mathbf{x} = \mathbf{0}\) is independent of time, which
is certainly the case for the regular sinusoid. Indeed we may obtain
\(\bar{\omega}\) under these assumptions, which we call
\emph{frequency matching}.  The result is shown in
\cref{fig:x-dependence}.

However, as we can see from \cref{fig:x-dependence}, the
frequency-matched solution (blue) and the numerical solution (orange),
will slowly drift apart, even though in the neighbourhood of
\(\mathbf{x} = \mathbf{0}\) and \(t=0\) the agreement is nearly
perfect. This ``drift'' is due to one of many differences between the
relativistic and Galilean simple harmonic oscillators. The presence of
\(\gamma(\dot{\mathbf{q}})\) means that \(\mathbf{q}\) and
\(\dot{\mathbf{q}}\) are not always exactly \(\pi/2\) out of phase,
which can be viewed as a change of effective mass of the mode leading
to a different frequency. We shall remove this drift from
\(\bar{\omega}\), and reabsorb it into \(\bar{\xi}\), as discussed in
the previous section (see also
Appendix~\ref{sec:integr-tagg-part}). Frequency-matching alone is
therefore not sufficient to produce an accurate analytical
description.

Instead, we must have two independent hypotheses for the forms of
\(\bar{\xi}\) and \(\bar{\omega}\) and fit them simultaneously. While
\(\bar{\omega}\) can itself be treated as a scalar, we need two more
parameters for \(\bar{\xi}\).

\begin{figure*} \centering
  \begin{subfigure}[t]{0.45\textwidth} \centering
    \includegraphics[width=\columnwidth]{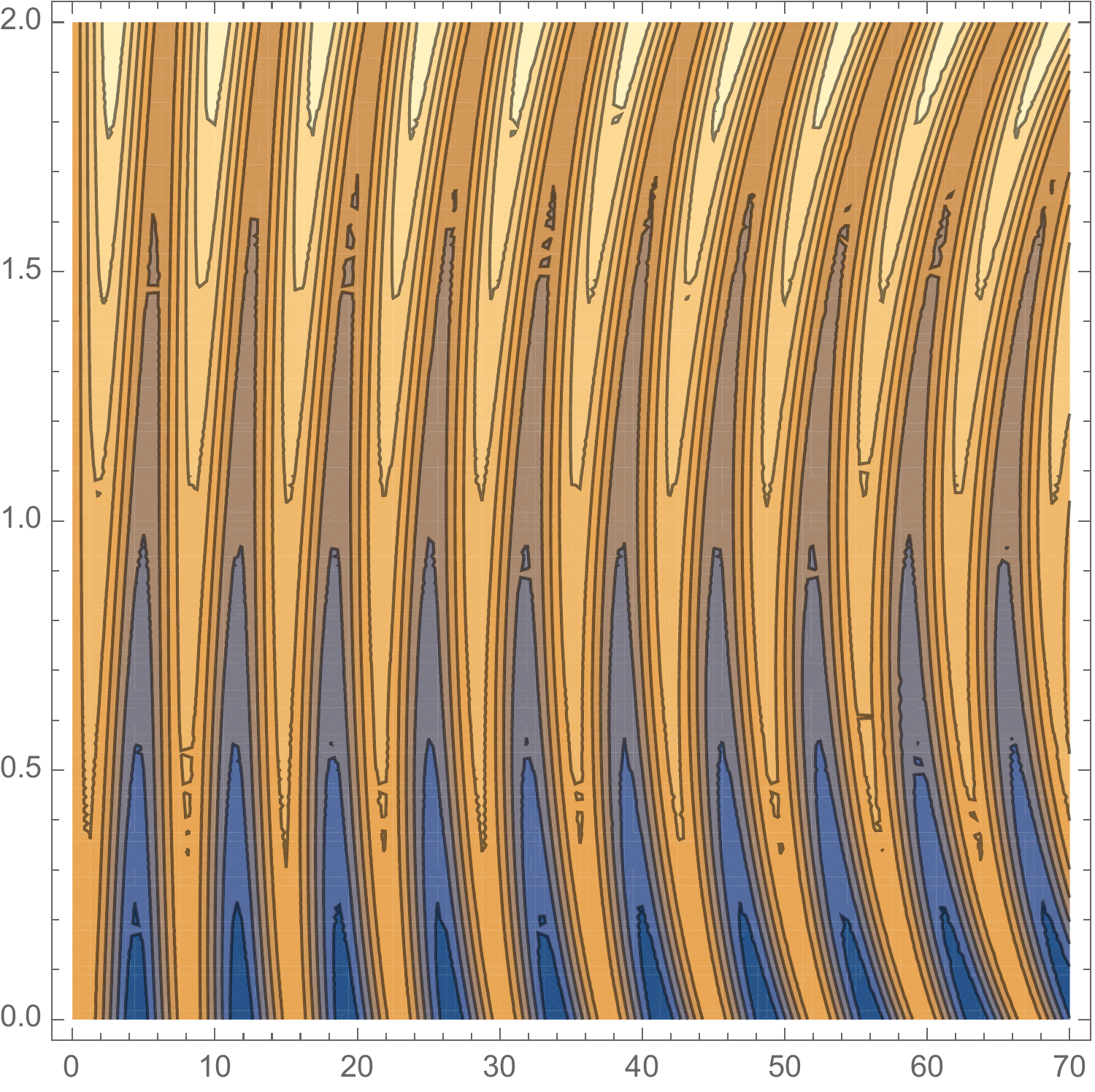}
    \caption{Strong coupling of \(g=1\). The increasing curvature of
      the sector where \(g x = 1\)  indicates either the
      \(\mathbf{x}\)-dependence of \(\bar{\omega}\) or the time
      -dependence of \(\bar{\xi}\). For reasons elaborated on in
      Sec.~\ref{sec:numeric-to-analytic}, we shall prefer the latter.}
  \end{subfigure}
  \begin{subfigure}[t]{0.45\textwidth} \centering
    \includegraphics[width=\columnwidth]{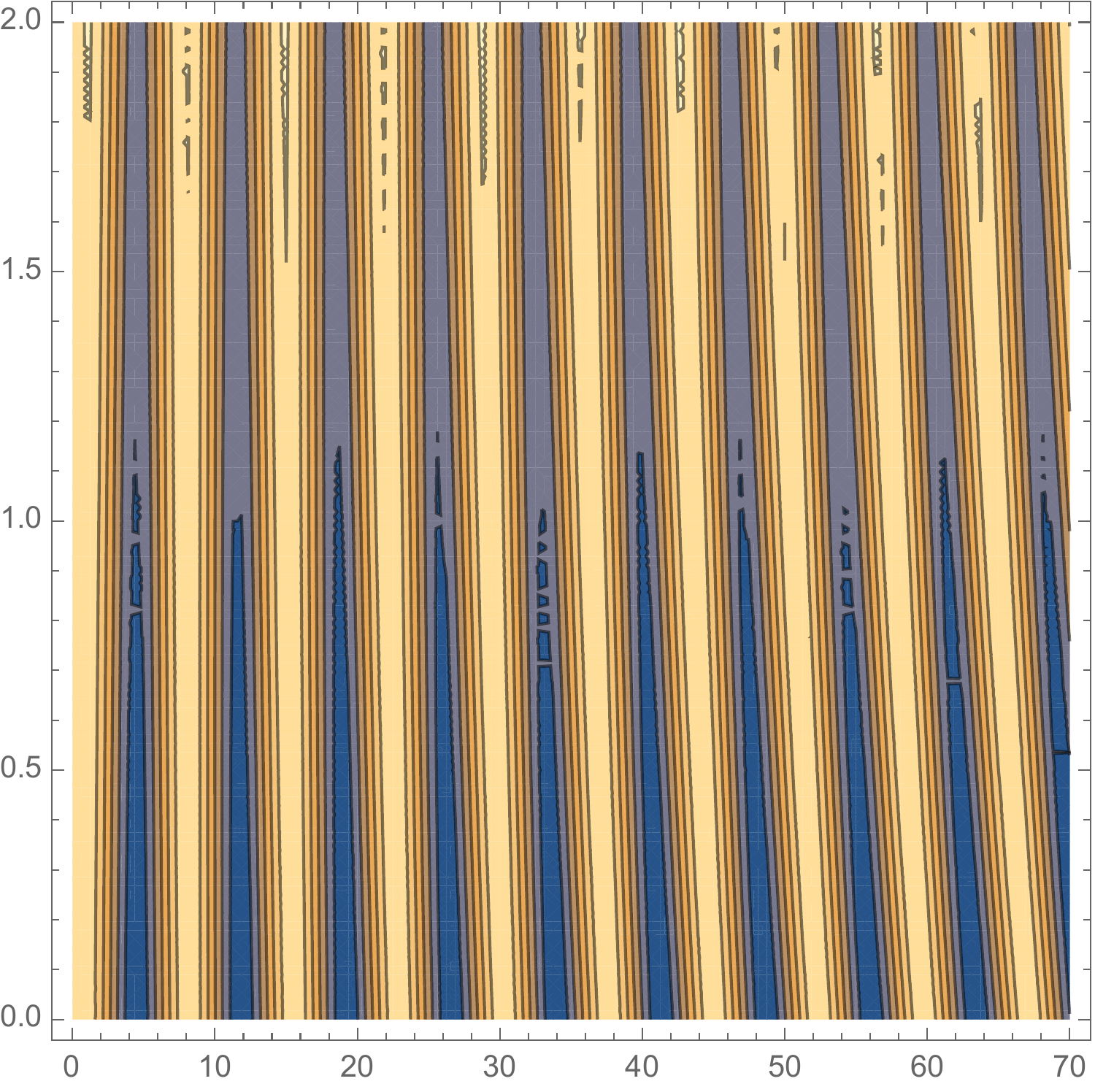}
    \caption{Weak coupling of \(g=0.2\). Note the near absence of
      \(\mathbf{x}\) dependence of the phase. This is because the
      relevant quantity is \(g x\) rather than \(x\), as one would
      expect from \cref{eq:N2-relat}. }
  \end{subfigure}
  \caption{Contour plots of the numerical solution \(q(x, t)\), where
    \(g\) is given in the sub-caption, and the remaining parameters
    (\(c, m, \omega\)) were set to \( 1 \) in their respective
    units. Color gradients represent the value of \(q(x, t)\) from
    small (blue) to large (orange). These plots illustrate the
    \(\mathbf{x}\) and \(t\) dependence introduced into the phase
    \(\bar{\xi}\) of the relativistic wave solution
    (\cref{eq:bath-motion-relat}).}\label{fig:xi-contour}
\end{figure*}

The shape in \cref{fig:xi-contour} can be approximated by an offset parabola:
\begin{equation}
  \bar{\xi}(t, x) = A t {(x - B)}^{2},
  \label{eq:xi-hypo}
\end{equation}
where \(A\) and \(B\) are the aforementioned parameters. Note that we
assume that \(\bar{\xi}(t= 0,x \neq 0) = 0\), which we need for
consistency with the initial conditions, but that
\({\bar{\xi}(t \neq 0, x = 0) \neq 0}\), and scales proportionally to
time, as we would expect in the first approximation to the drift we
saw in \cref{fig:x-dependence}.

By a least-square fit of the numerical solution using
\cref{eq:bath-motion-relat}, for the case of \(g = 1\) we obtain that
the best fit parameters are \(B=1.0000 \pm 0.0003\),
\(A = 0.0045 \pm 0.0005\) and \({\bar{\omega} = 0.78634 \pm
  0.0008}\).

Note that \(\bar{\omega}\) is not of the order of
  magnitude of \(\gamma\) for the relevant velocity. This shows that
  the change in frequency is not solely attributable to Lorentz time
  dilation.

\begin{figure}
  \centering%
  \includegraphics[width=\columnwidth]{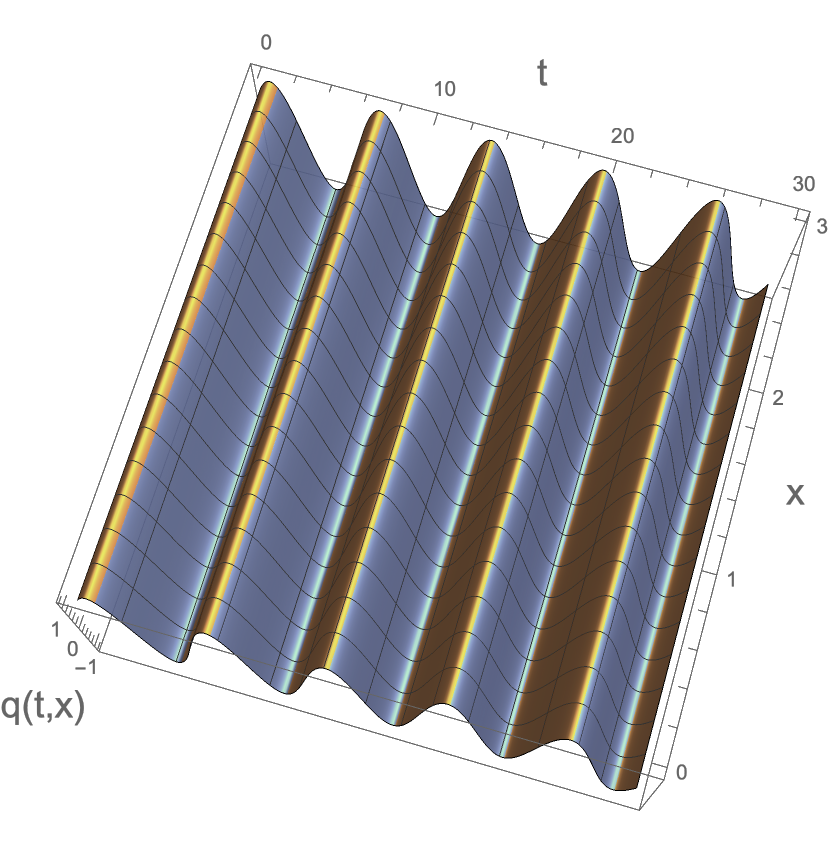}
  \caption{Comparison of the numerical solution to the heat bath
    equations of motion (orange) and the closed-form
    solution~\cref{eq:bath-motion-relat} (blue), after proper fitting
    of both \(\bar{\xi}\) and \(\bar{\omega}\).
    \(g = 0.2\).}\label{fig:corrected-match}
\end{figure}

Consequently, the results of the matching are presented in
\cref{fig:corrected-match}. Boundary conditions will require a
different hypothesis, but a similar approach, which leaves unchanged
the general form of the solution. Hence, with this general
prescription we can obtain an analytical form which is consistent with
the numerical solution, as demonstrated in \cref{fig:corrected-match}.

\subsection{Transverse component of the trajectory}\label{sec:transverse}
It is worth noting that the procedure can be similarly repeated for
\cref{eq:N2-relat-2}, which gives the transverse component of the acceleration
of the heat bath mode. The differential equation that defines the
equations of motion is different, by a factor of
\(\gamma^2(\dot{\mathbf{q}})\), which one may think should
significantly impact the applicability of \cref{eq:bath-motion-relat},
as an approximation. Our calculations show that the ansatz
\cref{eq:bath-motion-relat} is indeed a good approximation for this
equation as well, with different \(\bar{\omega}\) and
\(\bar{\xi}\). For the case of constant boundary conditions, that we
have discussed previously, the parabolic hypothesis for \(\bar{\xi}\)
with linear time scaling shows similar agreement, but with different
values of the parameters \(A\) and \(B\).

\section{Tagged particle dynamics}

Having justified the general structure of the solution for the bath
dynamics, based on the numerical results, we now proceed to
integrating by parts to obtain the dynamical trajectory of the tagged
particle using~\cref{eq:bath-motion-relat} calibrated as in the above
section. In general, the sinusoidal behavior in the integrand is not
independent of the trajectory \(\mathbf{x}(s)\) (recall that
\(\bar{\xi}(s) = \bar{\xi}(\mathbf{x}(s), s)\), where the implicit
dependence is the result of the parametrisation, and the explicit
dependence is the effect of absorbing the \(\mathbf{x}\) dependence
from \(\bar{\omega}\)), because of the dependence encoded in
\(\bar{\xi}(\mathbf{x}(t))\). From a mathematical point of view,
however, we can treat \(\bar{\xi}(\mathbf{x}(s))\) as ultimately a
generic function of \(s\), and write
\(\bar{\xi}(\mathbf{x}(s))\equiv \bar{\xi}(s)\), which is an equivalent
notation. This leads to the following integral for the last term on
the r.h.s.~of~\eqref{eq:bath-motion-relat}:
\begin{widetext}
  \begin{equation}
    \int_{0}^{t}\gamma(\mathbf{\dot{x}}(s))\mathbf{x}(s) \frac{\sin
      \bar{\omega}(t - \frac{\bar{\xi}(t)}{c} - s)}{\bar{\omega}} ds
    =
    \substitute{
      \int \gamma(\mathbf{\dot{x}}(s))\mathbf{x}(s) \frac{\sin \bar{\omega}(t -
        \frac{\bar{\xi}(s)}{c} - s)}{\bar{\omega}} ds }{t}\!\!- \substitute{\int\gamma(\mathbf{\dot{x}}(s))
      \mathbf{x}(s) \frac{\sin \bar{\omega}(t - \frac{\bar{\xi}(s)}{c} -
        s)}{\bar{\omega}} ds }{0}\!\! \label{eq:by-parts},
  \end{equation}
\end{widetext}
where, in the right hand side, we use the Newton-Leibniz formula for a
definite integral, expressed as two anti-derivatives. We have slightly
abused the notation here, in that the anti-derivative defines a set of
functions which differ up to a constant term, which one can in
principle determine from the boundary conditions. In the Galilean
case, the result would have been the same regardless of the boundary
conditions, but in the relativistic problem, we must determine the
integration constant, and only then substitute \(s\) with the proper
value. In this first step the necessity to find the integration
constant may not be immediately evident, but will become apparent as
we separate the terms.

We perform integration by parts in \cref{eq:by-parts}, and re-group
the resultant terms. Upon replacing in \cref{eq:bath-motion-relat},
and following the steps reported in
Appendix~\ref{sec:integr-tagg-part}, we finally get
\begin{widetext}
  \begin{align}\label{eq:bath-motion-relat-int}
    \begin{split}
      \mathbf{q}_{i}(t)\!-\!\frac{g_i\mathbf{x}(t)}{{(\bar{\omega}_i)}^2}
      = &\left[ \mathbf{q}_{i}(0) - \frac{ g_i}{\bar{\omega}_i^2}
        \mathbf{x}(0) \right] \cos \bar{\omega}_i \left( t -
        \frac{\bar{\xi}_{i}(t)}{c} \right)+ \gamma(\mathbf{\dot{q}}_i (0))\mathbf{\dot{q}}_i (0) \frac{\sin
        \bar{\omega}_i \left( t -\frac{\bar{\xi}_{i}(t)}{c} \right)}
      {\bar{\omega}_i} \\
      &+ \int_{0}^{t} \gamma
      (\mathbf{\dot{x}}(s)) \mathbf{\dot{x}}(s) \left\{
        \gamma^{-1}(\mathbf{\dot{x}}(s)) \int\gamma (\mathbf{\dot{x}}(s))
        \frac{\sin \bar{\omega}_{i}(t - \frac{\bar{\xi}(s)}{c} -s)}{\bar{\omega}_{i}} \,
        ds\right\} ds\\
      &+  g_i \frac{\mathbf{x}(0)}{\bar{\omega}_{i}^{2}} \substitute{\cos
        \bar{\omega}_i \left( t - \frac{\bar{\xi}_{i}(t)}{c} \right) + \int
        \gamma(\mathbf{\dot{x}}(s)) \bar{\omega}_{i} \sin \bar{\omega}_{i}(t -
        \frac{\bar{\xi}(s)}{c} -s) ds}{0}. \\
      &- g_i\frac{\mathbf{x}(t)}{\bar{\omega}_{i}^{2}}
      \substitute{
        \int \bar{\omega}_{i} \gamma(\mathbf{\dot{x}}(s)) \sin \bar{\omega}_{i}
        \left(t - \frac{\bar{\xi}(s)}{c} -s\right) ds - 1
      }{t} \\
    \end{split}
  \end{align}
\end{widetext}
This is the sought-after result, and the relativistic analogue
of~\cref{eq:bath-motion-classical-int}.

At this point it may be useful to reiterate the introduced
terms. Regarding the presence of \(\bar{\xi}_{i}\), for each heat bath
mode \(i\), this is connected to establishing initial conditions in a
statistical manner~\cite{zwanzig2001nonequilibrium}. In fact, we will
later assume that the initial conditions are drawn from a
Maxwell-Boltzmann-type distribution as in the Galilean
case~\cite{Zwanzig1973,zwanzig2001nonequilibrium}, which \emph{mutatis
  mutandis} is known as the J\"{u}ttner distribution in relativistic
statistical mechanics~\cite{Juettner,DunkelPRL}.  Upon moving to a
continuum of eigenfrequencies, the functions \(\bar{\xi}_{i}\) are
replaced by an extra parametric dependence on \(\bar{\omega}\). To
understand why, first recall how one moves to the continuum in the
Galilean case. The mapping between an index and the mode's frequency
is \emph{bijective}, by which we mean
\({\bar{\omega}_{i} \neq \bar{\omega}_{j}}\) \emph{if}, and
\emph{only} if \({i \neq j}\). So the index is interchangeable with
the frequency. Instead of stating that \(\bar{\xi}_{i}\) is the phase
function belonging to the \(i\)-th mode, we state that it is a
function \(\bar{\xi}(t, \mathbf{x} (t); \bar{\omega}_{i})\) that
depends on the frequency of the \(i\)-th mode. The summations with
respect to \(i\) are replaced with integrations with respect to the
continuum-spanning frequency variable \(\bar{\omega}\). The same is
true of all the parameters which vary mode-to-mode, picking up an
additional \(\bar{\omega}\) dependence.

Of course, here again we use the shorthand
\({\bar{\xi}(\bar{\omega}, t, \mathbf{x}(t)) = \bar{\xi}(\bar{\omega},
  t)}\), to make the notation more compact. The dependence on
\( \mathbf{x} \) is implicit and should be understood from the
context.

\section{Bayesian justification of the proposed analytical approximation}\label{sec:bayes-just-prop}
Formal verification of the results that form the basis of
Section~\ref{sec:numerics} is needed. One could have in principle
merely checked that substitutting \cref{eq:bath-motion-relat-int} into
\cref{eq:N2-relat,eq:N2-relat-2} for the appropriate parameters over a
certain domain of the free variables, e.~g.~\(\mathbf{x}\), produces
identity. Unfortunately, this forms a circular dependency, because the
values of \(\bar{\xi}\) and \(\bar{\omega}\) are established assuming
that the equations are correct to within machine precision.

Bayesian inference is widely used in statistical cosmology, particle
physics and gradually also in other branches of physics. It is widely
used for model comparison, particularly in cases where more
fine-grained analysis of multiple theories is necessary. In addition
to being able to reproduce any and all results obtained with
frequentist statistical methods, Bayesian inference is able to
automatically impose Occam's principle~\cite{occam} in model
comparison, thus avoiding ``overfitting'' (the famous ``Fermi elephant'').

We have based on \cref{eq:bath-motion-relat-int} implemented in Python
the following function:
\begin{align}
  \begin{split}
    \Delta = &\frac{ d^2 \mathbf{q}_i(t,\mathbf{x}; \bar{\xi}, \bar{\omega})}{dt^2}
      +  {\omega_i^2} \mathbf{q}_i(t,\mathbf{x}; \bar{\xi}, \bar{\omega})
      -  \mathbf{x},
    \label{eq:delta}
  \end{split}
\end{align}
where \(q_i\) is the approximate solution from \cref{eq:bath-motion-relat-int}, and \(\bar{\xi}\) is defined  in \cref{eq:xi-hypo}. Thus we have a hypothesis with three degrees of freedom: \(A\) and \(B\) from \cref{eq:xi-hypo} and \(\bar{\omega}\), and two nuisance parameters: \(x\) and \(t\).

For the likelihood we have chosen
\begin{equation}
    \mathcal{L}(A, B, \bar{\omega}) = e^{- \Delta^2}
\end{equation}
and the priors on the parameters were \(A \in (0, 100)\),
\(B \in (0, 2)\) and \(\bar{\omega} \in (0.001, 100)\), all uniform.
Inference with \texttt{PolyChord}~\cite{polychord} yielded an
uncorrelated Gaussian posterior with \(B=1.0000 \pm 0.0003\),
\(A = 0.0045 \pm 0.0005\) and \({\bar{\omega} = 0.78634 \pm 0.0008}\).
and a \emph{Bayesian log-evidence} of
\(\ln \mathcal{Z} = -18.5742 \pm 0.0005\). This result is to be
interpreted as follows: firstly, that our least-squares fit had been
accurate. Secondly, the evidence is within \(12\sigma\) of the result
we would expect if the parameters were truly normal-distributed ---
\(\ln \mathcal{Z}_\text{expected} = -18.5684\). It does not mean that
our analysis is invalid, quite the opposite, it suggests that the
parameters' joint distribution is not \emph{exactly} an uncorrelated
multivariate normal distribution, but that it is a very good
approximation: the difference is only \(0.03\%\) of the evidence.

One interprets the evidence by comparing it to a better, but much
more complex quasi-analytical solution to~\cref{eq:N2-relat-2} that can be obtained using Mathematica and that we do not report here as it is an extremely long expression (in the following referred to as ``quasi-analytical'' solution). We use
the same likelihood, but redefine \(\mathbf{q}_i\) in \cref{eq:delta}
to now be the quasi-analytical solution provided
by Mathematica~\cite{mathematica}. In this case the likelihood is independent of
\(A, B\) and \(\bar{\omega}\). As a result of being independent of two
parameters, the log-evidence jumps to
\(\ln \mathcal{Z} = -10.11 \pm 0.05\), which is both significantly
larger than the one we obtained earlier (due to the Occam penalty
associated with \(A, B\) and \(\bar{\omega}\)), but also indicative of
the fact that the quasi-analytical solution is not perfect. If
\(\Delta = 0\) for all values of \(t, x, A, B\) and \(\bar{\omega}\)
then the log-evidence is the log-volume of the prior space:
\(\ln \mathcal{Z} = -9.9\), which is the evidence one would have
obtained with the ``ground truth'' solution. This discrepancy of
\(3\%\) in the log-evidence tells us both that the quasi-analytical
solution is a reasonable approximation to the solution to
\cref{eq:N2-relat-2}. At the same time, however, it is not an exact analytical
solution, since \({(10.11 - 9.90) = 3\sigma}\). For comparison, if we
fix $A$, $B$ and $\bar{\omega}$, in our previous consideration, to their
best fit values and combine the errors in quadrature, we obtain
\(\ln \mathcal{Z} = -10.21 \pm 0.08\).

In quantitative terms, this means that the quasi-analytical
approximation is, with a \(e^{-10.11 - 9.9} = 81\%\) Bayes ratio, the
solution to \cref{eq:N2-relat-2}, and the approximation in
\cref{eq:bath-motion-relat-int} after parameter fitting --- is the solution with a Bayes ratio
\(74\%\). Larger Bayes ratios signal a better fit. Paraphrasing
Ref.~\cite{Trotta_2008}, we have found strong evidence for the
quasi-analytical approximation that one obtains from Mathematica, and marginally-strong evidence for the much 
more compact and manageable approximation in \cref{eq:bath-motion-relat-int}, provided that we do parameter fitting for
the boundary conditions specified in \cref{eq:bc-q,eq:bc-qdot}.
This analysis certainly justifies the trade-off between precision and compactness of the analytical expression in favour of \cref{eq:bath-motion-relat-int}, which will be used in the following to obtain a fully relativistic form of the Langevin equation.

\section{Relativistic Langevin Equation}\label{sec:relatv-lang-equat}
\subsection{Term classification}
To obtain the final relativistic Langevin equation, we substitute
\cref{eq:bath-motion-relat-int} into \cref{eq:eom-rel-x}. However, we
shall first analyze and classify the terms entering into the final
equation.

Analysing~\cref{eq:bath-motion-relat-int}, we can clearly notice
parallels to the original Galilean Langevin
equation~\cref{eq:langevin-classical}, as well as fundamental
differences. Let us start from the parallels.  There is a frictional
force term consisting of an integral, with the time derivative of the
coordinate \(\dot{\mathbf{x}}\), which multiplies the following memory
function:

\begin{subequations}\label{eq:defs-rel-discrete}
  \begin{widetext}
    \begin{equation}
      K'(t, s) = \sum_i \frac{g_i^2}{\bar{\omega}_i} \gamma^{-1}
      (\mathbf{\dot{x}}(s))
      \int \gamma (\mathbf{\dot{x}}(s))  \sin \bar{\omega}_i \left(t - s - \frac{\bar{\xi}_{i}(s)}{c}\right) ds.\label{eq:memory-relativistic}
    \end{equation}
    \end{widetext}
    This differs from the memory function of the Galilean case because of
    the renormalised frequency \(\bar{\omega} \) and because of the
    presence of the space-like parameter \( \bar{\xi}(t)\) needed to make
    the trigonometric functions manifestly Lorentz-covariant.

    Similarly to the Galilean case in Zwanzig's
    treatment~\cite{Zwanzig1973,zwanzig2001nonequilibrium}, one can
    identify terms that show dependencies on the boundary conditions, with
    the stochastic force:
    \begin{widetext}
    \begin{align}
      \begin{split}
        \mathbf{F}'_{p}(t)
        =
        \sum_i g_i \frac{m_i}{\gamma(\mathbf{\dot{x}}(t))} 
        \Bigg\{
        &\left[
          \mathbf{q}_{i}(0)- \frac{ g_i}{\bar{\omega}_i^2} \mathbf{x}(0)
        \right]
        \cos \bar{\omega}_i\left(t - \frac{\bar{\xi}_{i}}{c} \right)
        + \gamma(\mathbf{\dot{q}}_{i}(0))\mathbf{\dot{q}}_i (0) \frac{
          \sin \bar{\omega}_i \left( t -\frac{\bar{\xi}_{i}(t)}{c} \right)
        }{\bar{\omega}_i}\\
        &+ g_i \frac{\mathbf{x}(0)}{\bar{\omega}_{i}^{2}} \substitute{\cos
          \bar{\omega}_i \left( t - \frac{\bar{\xi}_{i}(t)}{c} \right) + \int
          \gamma(\mathbf{\dot{x}}(s)) \bar{\omega}_{i} \sin \bar{\omega}_{i}(t -
          \frac{\bar{\xi}(s)}{c} -s) ds}{0}
        \Bigg\}.\label{eq:stochastic-force-relativistic}
      \end{split}
    \end{align}
  \end{widetext}
  Here, exactly like in the original Galilean derivation
  by~\citeauthor{Zwanzig1973}~\cite{Zwanzig1973}, the stochasticity of
  the force follows from ``ignorance'' about the boundary
  conditions~\cite{zwanzig2001nonequilibrium}. Had we not
  coarse-grained away the information about the full \emph{microstate}
  of the system, the stochastic force would appear fully
  deterministic~\cite{zwanzig2001nonequilibrium}.  It can be shown (in
  Section~\ref{sec:relat-fluct-diss}) that the above force term
  \(\mathbf{F}'_{p}\) given by
  \cref{eq:stochastic-force-relativistic} has zero average,
  \(\langle {F}'_{p}(t) \rangle = 0\), as required for the stochastic
  force in the Langevin equation~\cite{zwanzig2001nonequilibrium}. We
  shall discuss this in more detail in subsequent sections.

  However, there are new terms that are not present in the Galilean
  equation (if \( \bar{\xi}_{i}(t, \mathbf{x}) \) were to be treated
  as a small parameter, then all such terms would be
  \( O(\bar{\xi}) \)), such as:
  \begin{widetext}
    \begin{equation}
      \mathbf{F'}_{r} (t)
      =
      - \sum_{i}\frac{m_{i} g^{2}_i}{\gamma(\mathbf{\dot{x}}(t))}
      \frac{\mathbf{x}(t)}{\bar{\omega}_{i}^{2}} \substitute{
        \int \bar{\omega}_{i}
        \gamma(\mathbf{\dot{x}}(s)) \sin \bar{\omega}_{i}\left(t - \frac{\bar{\xi}(s)}{c} -s\right)ds
        - 1
      }{t}.
      \label{eq:restoring-force}
    \end{equation}
  \end{widetext}
\end{subequations}
Interestingly, this looks like a restoring force where the ``spring
constant'' is dependent on the trajectory through
\(\bar{\xi}_i(\mathbf{x}(s))\) --- contained in the r.h.s.~of the
above equation --, which affects the overall dynamics of the
system. In other words, \cref{eq:restoring-force} describes a
restoring force which retains ``memory'' of the dynamics, in a similar
way as the memory encoded in the stochastic force. To our knowledge,
this new, \emph{emergent} ``restoring force'' induced by the coupling
to the bath in the fully-relativistic regime has never been derived or
discussed in previous literature.

Thus the final equation takes the form

  \begin{equation}
    \label{eq:langevin-non-symmetric}
    \frac{d \left[ \gamma(\mathbf{\dot{x}}) m \mathbf{\dot{x}}\right]}{dt}\!
    =
    \!\mathbf{F'}\!\!\!_{p}
    +  \mathbf{F}_{\text{ext}}\!
    +  \mathbf{F'}\!\!\!_{r}
    -\!\int_0^t\!\!\!\gamma(\mathbf{\dot{x}}(t-s))\dot{\mathbf{x}}(t\!-\!s) K'(t,s)\, ds,
  \end{equation}

where we use the prime to indicate that the primed terms are different
in the relativistic formulation compared to the Galilean case.  This
equation is the central result of this paper, and represents a
full-fledged relativistic Langevin equation. To our knowledge, this is
the most general form of relativistic Langevin equation proposed so
far, and recovers the previously known forms of Langevin equations
(non-relativistic Generalised Langevin equation, Langevin eq.~for
relativistic tagged particle in non-relativistic bath, and weakly
relativistic Langevin eq.) in the relevant limits.

In the next section we further discuss the structure of this equation
and the significance of its terms, as well as the various limits that
can be recovered. We will also address aspects related to symmetry
implications connected with~\cref{eq:langevin-non-symmetric}.

\subsection{\(\bar{\xi}\) and \(\bar{\omega}\) in practice}

It must be noted, that so far we have not used any specific value or functional form for either \(\bar{\omega}_{i}\) or \(\bar{\xi}_{i}\). Indeed, the two are ``integration constants'' in the sense that their value is fully determined by initial conditions. We have shown \emph{a} functional form that can be obtained for a very specific set of initial conditions in Sec.~\ref{sec:matching}, and while these specific initial conditions (\cref{eq:bc}) are chosen at random, the algorithm of determining \(\bar{\xi}\) and \(\bar{\omega}\) should be sufficiently general. 
    
One can impose the initial conditions based on a specific model of interaction. For example if the heat bath comprises of a mass on a spring that enters elastic contact interaction with the tagged particle, at equilibrium, then the initial conditions are straightforward: 
    \begin{subequations}\label{eq:bcc}
        \begin{align}
            \mathbf{q}(0, \mathbf{x}) = &0 \label{eq:bcc-q}\\
            \dot{\mathbf{q}}(0, \mathbf{x}) = &-1, \label{eq:bcc-qdot}
        \end{align}
    \end{subequations}
    While this particular situation is highly unlikely to occur in real systems, the principle can be generalised.

\section{Discussion}

\subsection{Stochasticity}\label{sec:stochasticity-of-eqn}
We start our discussion of the Lorentz-covariant Generalised Langevin equation by noting that much like its Newtonian counterpart
  \cref{eq:langevin-classical}, it too is fundamentally deterministic.
  While some variants of the generalized Langevin equation share this
  property, not all versions are indeed deterministic or
  time-reversible.  While what we have found is indeed a rather
  general statement about the behaviour of many systems, the true (and
  ultimate) derivation of the generalized Langevin equation must allow
  for intrinsic non-determinism in the relativistic limit.

  In the Newtonian case, the effective non-determinism stems from two
  main arguments: the first is that the heat bath is too complex for
  the difference between deterministic motion and chaotic behaviour to
  be measurable. The same is true in special relativity, more so
  because of the issues of relative simultaneity.

  The second argument notes that, in the limit of Brownian motion, the tagged particle is often much
  heavier than the constituents of its environment and thus the
  time-scales for the measurable behaviour of the Brownian particle
  are much longer than the time-scales needed to measure changes in
  the environment. In effect, the heat bath moves too quickly for its
  behaviour in the time scale of the tagged particle to be anything
  but random noise. This argument is less obviously true in special
  relativity: the faster moving environmental particles experience
  time more slowly than the tagged particle. However these
  complications vanish once one affixes the lab frame to the
  instantaneous rest frame of the tagged particle. In this frame of
  reference, the particles' subjective experience of time is
  irrelevant to the statistics of their motion.

  Thus we fully expect the relativistic generalized Langevin equation
  to behave as ``effectively stochastic'' in many circumstances. It is
  possible, that the circumstances differ from the ones in which the
  Newtonian equation would have been effectively stochastic, which
  would require further investigation in future work.

\subsection{Lorentz covariance}\label{sec:lorentz-covariance}
We are now equipped to consider the Lorentz covariance of the Langevin
equation. Our first task is to show that the starting particle-bath
Lagrangian is indeed Lorentz-covariant.  This is tantamount to show
that the action corresponding to \cref{eq:starting-lagrangian-rel} is
Lorentz invariant: a scalar, a vector, etc. We thus consider the
action associated with our starting particle-bath Lagrangian:
\begin{align}
  \begin{split}
    S
    =
    &\int\!\! d t ~\,
    \frac{mc^{2}} {\gamma(\dot{\mathbf{x}})}
    - V(\mathbf{x}, \mathbf{\dot{x}}, t)
    + \sum_i \frac{m_{i}c^{2}}{\gamma(\dot{\mathbf{q}}_{i})}\\
    &+ \! \sum_{i}\frac{ m_{i}\omega^{2}_{i}}{2\gamma(\mathbf{\dot{x}})}\!
    \left[
      c^2{
        \left(
          t_i - \frac{g_i}{\omega_{i}^{2}} t
        \right)}^2\!\!\!
      - {\left(
          \mathbf{q}_i - \frac{g_i}{\omega^{2}_{i}} \mathbf{x}
        \right)}^{2}
    \right]\!\!.
  \end{split}
\end{align}
The integral decouples into a series of terms: the kinetic terms of
the bath eigenmodes and the tagged particle, the external Lorentz
force acting on the external particle, and the interaction between
bath eigenmodes and the tagged particle.

The kinetic terms simplify to integrals with respect to proper time of
Lorentz scalars, e.g.
\begin{align}
  \label{eq:kinetic-covariance}
  \int dt \, \frac{mc^{2}}{\gamma(\mathbf{\dot{x}(t)})} = \int d \tau \, mc^{2},
\end{align}
and similarly for the bath eigenmodes. The external force acts
precisely like the electromagnetic force (which historically was the
first Lorentz covariant interaction discovered), meaning that the
\(V(\mathbf{x}, \mathbf{\dot{x}}, t)\) term is also Lorentz covariant.

Finally, since we model our interaction as a propagation of
null-separated events, the interval, regardless of pre-factors would
be a Lorentz scalar --- zero. However, had we not restricted ourselves
to null-separated events, the integral would become
\begin{align}
  \label{eq:interaction-covariance}
  \sum_{i} \frac{1}{2}  \int d\tau \, m_{i} \omega_{i}^{2}c^{2}\delta \tau_{i}^{2},
\end{align}
where in this case the \(\delta \tau_{i}\) is the Lorentz invariant
proper time interval separating the interaction events. This being the
final term in the starting Lagrangian proves that we have constructed
a Lorentz-covariant theory of the interactions.  It is well-known that
the equations of motion which correspond to a Lorentz covariant
Lagrangian, are themselves automatically
Lorentz-covariant~\cite{cahill_2019}.

However, the Langevin equation~\eqref{eq:bath-motion-relat} from which
we obtain~\eqref{eq:langevin-non-symmetric} is an approximate solution
to the equations of motion~\eqref{eq:eom-rel-x}. Thus we must also
demonstrate that \cref{eq:bath-motion-relat} and the resultant
\cref{eq:langevin-non-symmetric} is Lorentz covariant.  In
practice, since the Langevin equation is a covariant equation for
non-relativistic 3-vectors, what we must demonstrate is that the
components on each side transform as spatial components of a Lorentz
4-vector.

For \cref{eq:bath-motion-relat}, after a boost with velocity
\(\mathbf{u}\) in some direction from the initial inertial reference
frame \(S\) into the reference frame \(S'\), only the components of
\(\mathbf{q}\) along the direction \(\mathbf{u}\) are affected. All
trigonometric functions transform in the same way, and consistently,
so that the equation holds for \(\mathbf{q}'(t')\) in \(S'\). Namely,
the \(\bar{\omega}\) transforms as the inverse of a time-scale, \(t\)
on both sides transforms as a time-scale. If we left it at that, the
right-hand-side phase would differ from the corresponding phase on the
left-hand-side. Fortunately, we can absorb the difference between the
two into the laws of transformation of \(\bar{\xi}\), as we already
did when introducing this parameter. What remains is a length at the
initial time \(\mathbf{q}(0)\), a 4-velocity multiplying a time scale
\(1 / \bar{\omega}\), and an integral. The integrand transforms like a
length multiplied by two time-scales: \(s\) and \(1 / \bar{\omega}\),
which we ``fix'' by noting that \(g_i\) must transform like the
inverse of the square of a time-scale. Consequently, the left-hand
side and the right hand side transform as three-components of Lorentz
four-vectors.

For \cref{eq:langevin-non-symmetric}, much like in the previous case,
let's proceed in a term-by-term fashion. Firstly, the term
\begin{align}
    \frac{d}{dt}
    \left[ \gamma(\mathbf{\dot{x}}) m \mathbf{\dot{x}}\right]
    =
    \frac{d}{d t} p^{i}
\end{align}
is a standard~\cite[Eq.~12]{dunkel_hanggi} Lorentz-covariant inertial
term. Adding the fourth (temporal) component~\footnote{In covariant
  notation, Greek indices denote four-vectors, while Latin indices ---
  spatial three-vectors} --- \(E/c\) this becomes the tagged
particle's four momentum.  Thus, in order to demonstrate that
\cref{eq:langevin-non-symmetric} is compatible with special
relativity, we must show that the right-hand side also transforms as a
4-vector of the same kind as the left-hand side~\footnote{The contra-
  and co- variant vectors on the left-hand side must transform into
  contra- and co- variant vectors on the right-hand side. }.  The
three forces, \(\mathbf{F}'_p, \mathbf{F}_\text{ext}\) and
\(\mathbf{F}'_r\) are spatial components of contravariant 4-vectors,
which we imposed by construction, with the temporal components of the
4-forces being equal to zero. Thus, the only term whose Lorentz
covariance must be demonstrated is the memory term.  Specifically it
must transform as a (contravariant) 4-vector.

We shall adopt the same approach as~\cite{dunkel_hanggi}. Namely, we
shall introduce the memory tensor \(K^{\mu}{}^{\nu}\), and re-write
the integral as
\begin{align}
  \int ds\, K^\mu{}_\nu (t, s) \frac{p^\nu (t-s)}{m}
\end{align}
with implicit sums over repeated indices (\(\mu\)). This is a
contravariant 4-vector, provided that the memory tensor transforms as
a Lorentz covariant rank-2 tensor, which we again, impose by
construction following~\cite{dunkel_hanggi}. Specifically it's a
diagonal matrix, with null temporal components, and diagonal entries
all equal to~\cite[See Eq.~15]{dunkel_hanggi} \(K'(t)\), defined in
\cref{eq:memory-relativistic}. This makes
\cref{eq:langevin-non-symmetric} a direct equivalent of Eq.~(16)
in~\cite{dunkel_hanggi}.

Thus, we showed that all terms appearing
in~\cref{eq:langevin-non-symmetric} are Lorentz-covariant
4-vectors. This allows us to rewrite our relativistic Langevin
equation~\eqref{eq:langevin-non-symmetric} in explicitly
Lorentz-covariant form:
\begin{equation}
  \frac{d}{d t} p^{\mu} 
  =
  F'^{\mu}_{p}
  + F^{\mu}_{\text{ext}}
  + F'^{\mu}_{r}
  - \int\! ds\, K^{\mu}{}_\nu (t, s) \frac{p^\nu (t-s)}{m}.
  \label{eq:covariant-langevin}
\end{equation}
where we still used coordinate time $t$ since we work in the instantaneous rest frame
of the tagged particle, hence we do not have a difference between proper and coordinate time. Hence, we could equivalently replace $t$ with $\tau$ in the above equation. 

\subsection{Slow (non-relativistic) heat bath limit}

Before we continue, it's important to consider whether
\cref{eq:langevin-non-symmetric} is compatible with the Langevin
dynamics as is already known from previous works, by studying the
relevant limits.

The first step is to analyze how \cref{eq:langevin-non-symmetric}
differs from the fully-Galilean limit,
i.e. \cref{eq:langevin-classical}. There are three key differences:
(i) a \( \gamma \) factor in the acceleration term, (ii) differences
in how the stochastic force and the friction kernel are defined, and
(iii) an extra term \( F_{r}' \) that is proportional to
\( \mathbf{x}(t) \). The first difference is trivial, in the
non-relativistic (Newtonian) physics, \( \gamma \rightarrow 1 \). The
primed version of the friction kernel, and the stochastic force,
\cref{eq:memory-relativistic} and
\cref{eq:stochastic-force-relativistic}, differ from
\cref{eq:memory-classical} and \cref{eq:noise-classical}, in a
peculiar way. If we were to treat
\( \bar{\xi}_{i}(t, \mathbf{x}(t)) \) as a small parameter, and expand
the relevant equations, we would see that all the differences between
the non-relativistic terms of the Langevin equation and the
relativistic counterparts derived above are of order
\( O(\bar{\xi}_{i}) \).

That is, we have verified that in the Galilean non-relativistic
regime, \( \bar{\xi}_{i} \rightarrow 0\) and
\(\bar{\omega} \rightarrow \omega\), upon taking the limit of the slow
heat bath, i.e.  \( \gamma(\dot{\mathbf{q}}) \rightarrow 1 \).  Hence,
upon further setting \(\gamma =1\) in the acceleration term, the
Galilean Langevin equation is correctly recovered in the appropriate
limits by our \cref{eq:langevin-non-symmetric}.

Furthermore, starting from the general relativistic
\cref{eq:langevin-non-symmetric}, one can choose to reduce the
equations of motion of the bath \cref{eq:eom-rel-q} to their Galilean
counterpart \cref{eq:heat-bath-momentum}, while not taking the limit
\( \gamma(\mathbf{\dot{x}}) \rightarrow 1 \). Therefore the Langevin
equation reduces to
\begin{equation}\label{eq:lang-relat}
  \frac{d}{dt} \left[
    \gamma(\mathbf{\dot{x}}) m \mathbf{\dot{x}}
  \right]\!
  =
  \!\mathbf{F_p}\!
  -\! \mathbf{F}_\text{ext}(t, \mathbf{x}, \mathbf{\dot{x}})\!
  -\!\!\int_0^t\!\!\! \gamma(\mathbf{\dot{x}}(s))\dot{\mathbf{x}}(t\!-\!s)K'(t, s)\, ds,
\end{equation}
where we notice that all the force terms are almost exactly the same
as in the Galilean Langevin equation, and the term
\( \mathbf{F}'_r \rightarrow 0 \).  However, \(\gamma\) can still be
\(\gg 1\), implying that the tagged particle moves at
relativistic speeds in an otherwise non-relativistic bath.  This limit
given by \cref{eq:lang-relat}, is equivalent to the equations derived
by \citet{Debbasch,Quasirel-langev-eq}, which describe the motion of a
relativistic tagged particle embedded in a heat bath of
non-relativistic oscillators.

\subsection{Symmetries}\label{sec:symmetry}

During our derivation we have pointed out the necessity of letting
\( \bar{\xi}_{i}(t, \mathbf{x}(t)) \) be a function of time and of the
trajectory of the tagged particle. This manifests as breaking of the
following symmetries. If
\( \tfrac{\partial \bar{\xi}} {\partial t} \neq 0\), we no longer deal
with harmonic waves, and the propagation is no longer exactly
periodic, thus violating time-translation invariance. For anything but
the weakest coupling,
\( \tfrac{\partial \bar{\xi}} {\partial \mathbf{x}} \neq 0\), if we
were to shift the origins of both \( \mathbf{q} \) and
\( \mathbf{x} \), we wouldn't necessarily recover the same
equation. Similarly, the inversions of time \( t \), or of space, for
\( \mathbf{q} \) and \( \mathbf{x} \), would not recover the same
equation. Nonetheless, the simultaneous inversion of both, would leave
the equation unchanged if we were to extend the definition
\(\bar{\xi}_i(-t, - \mathbf{x}) = - \bar{\xi}_i(t, \mathbf{x})\). In
other words we have lost parity covariance and time translation
invariance.

However, as we shall see further below in this section, the loss of
inversion symmetries is a natural consequence of combining statistical
mechanics with special relativity, and the translation invariance
isn't lost in the strict sense, just it doesn't appear to affect the
length-like parameters of the problem.

\subsubsection{Time inversion and parity}\label{sec:time-invers-parity}

Let's first address the less controversial of the two apparently
``lost'' symmetries.  Time and space inversion symmetry are violated
individually. However, since \cref{eq:eom-rel-q} is invariant with
respect to a simultaneous inversion of both, and the function which
approximates the general solution (\cref{eq:bath-motion-relat}), is
too, we can expect \cref{eq:langevin-non-symmetric} to be invariant
with respect to inversion of both time \emph{and} space, combined
together.

This is not uncommon: CPT~\cite{Y98thestatus,CPT-Schwinger} is widely
considered to be a fundamental symmetry of the relativistic theories
while symmetries with respect to time and space inversion are
individually violated even in simple Newtonian cases: a disk spinning
clockwise is spinning counter-clockwise in the mirror universe, and a
counter-clockwise spinning disk is spinning clockwise if we reverse
the arrow of time.

On the larger scales, one also has the thermodynamic arrow of time,
which reflects the fact that irreversible processes flow in the
direction of increasing entropy. This is a consequence of the breaking of
time inversion invariance which is already at play in the Langevin
equation~\cite{zwanzig2001nonequilibrium}. However, we also predict a
loss of parity invariance, which is consistent with recent
observations~\cite{Minami_2020}. It is yet to be explored at this
point if the scale or extent of parity violations is similar for the
experimental observations and our theory, but if it is, this might
suggest that the thermodynamic arrow of time induced by dissipation
leads to a thermodynamic chirality of the structures in the universe,
that may have imprinted early in the universe's development, when the relativistic Langevin
equation more accurately reflected its behavior.

\subsubsection{Space-time translation invariance}\label{sec:space-time-transl}
In this section we shall briefly address the apparent breaking of
translational invariance in \cref{eq:langevin-non-symmetric}. Let us
first recall what kind of translational invariance is relevant
here. Let us rescale the variables
\begin{subequations}
    \begin{align}
  \mathbf{Q}_{i} = &\mathbf{q}_{i} \frac{\omega_{i}^{2}}{g_{i}}\\
  \Omega_{i}^{2} = &g_{i}\\
  M_{i} = &m_{i} \frac{g_{i}}{\omega_{i}^{2}}.
\end{align}
\end{subequations}
Then we substitute into \cref{eq:c-l-classical}, to yield
\begin{equation}
  L
  =
  \frac{m\mathbf{\dot{x}}^{2}}{2}
  + \sum_{i} \frac{M_{i}\mathbf{\dot{Q}}_{i}^{2}}{2}
  - V(\mathbf{x})
  - \frac{M_{i}\Omega_{i}^{2}}{2}
  {\left( \mathbf{Q}_{i} - \mathbf{x}\right)}^2,
  \label{eq:c-l-length}
\end{equation}
see also Ref.~\cite{Ambegaokar}.  This Lagrangian is invariant for
\(V(\mathbf{x})=0\) under global translations of both the tagged
particle and the bath, as is well known for the Caldeira-Leggett
model~\cite{Ambegaokar,Weiss}.  Its equations of motion and the
respective solutions, are left unchanged if the origins of
\( \mathbf{x} \) and \( \mathbf{Q} \) are both displaced by the same
arbitrary vector \( \mathbf{A} \in R^{3} \).

The same cannot be said of \cref{eq:langevin-non-symmetric}, primarily
because \(\bar{\xi}_{i}\) depends on \( \mathbf{x} \), and the
displacement of the origin of \( \mathbf{x} \) would then lead to a
change in the equation.

It is important to recall that, in the original Caldeira-Leggett
Lagrangian or in general, we do not have enough terms to produce a
complete square as in \cref{eq:c-l-length}. The missing counter-terms
are proportional to \( \mathbf{x}^{2} \), and \emph{a priori} we have
no physical interaction that would provide them, apart from the
special choice of coupling constants that leads to
\cref{eq:c-l-length} shown above. So, as is common
practice~\cite{Weiss,Ambegaokar}, rather than assuming a physical
model for the interaction, we have imposed the translational
invariance on the Lagrangian, and effectively swept the difference
between the true Lagrangian corresponding to our model and the one
that is translationally invariant into \( V(\mathbf{x}) \).  Thus,
translational invariance is not a natural symmetry of the
Caldeira-Legett model although it can be recovered for a suitable
choice of coupling constants~\cite{Ambegaokar}.

While the trick of adding counter-terms gives the illusion of
translational invariance, it does not in fact resolve a real
problem. We could have, by virtue of including a vector potential in
\cref{eq:starting-lagrangian-rel} renormalised the terms to not
contain any apparent violations of translational invariance, but that
would obscure the fact that the spatial coordinates are not, in fact,
coordinates. It should be clear that in
\cref{eq:langevin-non-symmetric} there is actually no real breaking of
translational invariance with respect to coordinate reparametrization:
\( \mathbf{x} \) really is a displacement from a physically
significant location, hence both \( \mathbf{x} \) and
\( \mathbf{q}_{i} \) should individually be invariant with respect to
changes of coordinate origins.  This consideration becomes even more
relevant if one considers that, as will be elaborated on below in
Sec.~\ref{sec:continuum}, the coordinates \( \mathbf{x} \) (and to
some extent also \( \mathbf{q}_{i} \)) are meant to signify
``displacements'' from an initial thermodynamic state.

\subsection{Continuous spectrum of the bath eigenmodes}\label{sec:continuum}
In this section we shall attempt to provide a foundation for the
relativistic extension of a cornerstone of nonequilibrium statistical
mechanics, the fluctuation-dissipation
theorem~\cite{Koide2011,Pal_2020}. We compare the predictions of the
Galilean and special relativity principles in the context of
statistical mechanics, for simple models.

We start by focusing on the definitions of the stochastic force and
memory function:~\cref{eq:memory-classical}
and~\cref{eq:noise-classical}. The two sums can be regarded as Fourier
transforms if one replaces \(\sum_{i}\) with an
integral\(\int_0^\infty \rho(\bar{\omega}) d\bar{\omega}\), by
introducing a density of states \(\rho(\bar{\omega})\), and promoting
\(\bar{\omega}_{i}\) to a continuous variable.  By analogy, we should
expect there to be Fourier representations of the memory function, of
the restoring force and of the stochastic force. Namely,
\begin{widetext}
\begin{subequations}\label{eq:defs-rel-cont}
    \begin{align}
      K'(t, s) = \!\int_{0}^{\infty}\!\!d\bar{\omega}\,
      \rho(\bar{\omega}) \frac{
      g^2(\bar{\omega})}{\bar{\omega}} \gamma^{-1}(\mathbf{\dot{x}}(s))\int\gamma (\mathbf{\dot{x}}(s))  \sin \bar{\omega} \left(t - s -
      \frac{\bar{\xi}(\bar{\omega}, s)}{c}\right) ds,\label{eq:rel-memory-integral}
    \end{align}
    \begin{align}
        \begin{split}
          \mathbf{F'}\!\!\!_{p}(t)
          =
          \!\int_{0}^{\infty}\!\!\!d\bar{\omega}
          \, \rho(\bar{\omega}) g(\bar{\omega})
          \frac{m(\bar{\omega})}{\gamma (\mathbf{\dot{x}}(t))}
          \Bigg\{
            &\left[\mathbf{q}(\bar{\omega};0)-\frac{g(\bar{\omega})}{\bar{\omega}^2}\mathbf{x}(0) \!\right]\cos\bar{\omega}\!\left(t\!-\!\frac{\bar{\xi}(\bar{\omega}, t)}{c}\!\right)
            + \gamma(\mathbf{\dot{q}}(0)) \mathbf{\dot{q}}_i (0) \frac{\sin \bar{\omega} \left( t
                -\frac{\bar{\xi}(\bar{\omega}, t)}{c} \right)}{\bar{\omega}}\\
            &+g(\bar{\omega}) \frac{\mathbf{x}(0)}{\bar{\omega}^{2}} \substitute{\cos
              \bar{\omega} \left( t - \frac{\bar{\xi}(\bar{\omega}, t)}{c} \right) + \int
              \gamma(\mathbf{\dot{x}}(s)) \bar{\omega} \sin \bar{\omega}(t -
              \frac{\bar{\xi}(\bar{\omega}, s)}{c} -s) ds}{0}
            \Bigg\}, \label{eq:stochastic-force-relativistic-integral}
        \end{split}
    \end{align}

    \begin{align}
      \mathbf{F'}\!\!\!_{r}(t)
      =
      - \int_{0}^{\infty}\!
      d\bar{\omega}\,\rho(\bar{\omega}) \frac{m(\bar{\omega}) g^{2}(\bar{\omega})}{\gamma(\mathbf{\dot{x}}(t))} \frac{\mathbf{x}(t)}{\bar{\omega}^{2}}
      \substitute{\int \bar{\omega} \gamma(\mathbf{\dot{x}}(s)) \sin \bar{\omega}\left(t - \frac{\bar{\xi}(\bar{\omega},s)}{c} -s\right)ds - 1}{t}. \label{eq:restoring-force-integral}
    \end{align}
  \end{subequations}
which completes our derivation.
\end{widetext}

Of note is the following difference between~\cref{eq:defs-rel-cont}
compared to their discrete counterparts~\cref{eq:defs-rel-discrete}:
the indices \(i\) are all replaced with a parametric dependence on
\(\bar{\omega}\) by suitably introducing a density of states of the
bosonic bath vibrations~\cite{zwanzig2001nonequilibrium}. However, as
in the discrete case, no residual dependence on frequencies is
observed also in the continuum case: everything which depends on
\(\bar{\omega}\) is integrated out.

\subsubsection{Markovian limit}
Of particular interest is the case wherein
\({\rho(\omega) = \alpha \omega^{2}}\), as for bosonic particles, and
\({g(\omega) = \text{Const}}\). If applied in the limit of low
velocities, ignoring all relativistic \(O(\bar{\xi})\) effects, the
memory function's integral reduces to an integral from zero to
infinite of a simple $\cos \bar{\omega}_{i} t$, thus
\({K'(t) = K(t) \propto \delta(t)}\), as shown
in~\cite{zwanzig2001nonequilibrium}. Therefore, the entire Langevin
equation becomes Markovian, in the sense that the viscous response at
time \(t\) depends on the velocity \(\dot{\mathbf{x}}(t)\) evaluated
at the same
time \(t\), and there are no memory effects.

In the relativistic case, we do not have a simple cosine function
inside the integral in \cref{eq:rel-memory-integral}, but an integral
which also contains the dependence on the bath oscillator trajectory
via $\bar\xi$.  Upon inserting the form~\eqref{eq:xi-hypo} that we
obtain from numerics-assisted parametrization of trajectories, the
inner integral in \cref{eq:rel-memory-integral} can be evaluated
analytically, and gives an expression of the type
\({\mathcal{A}(x)}^{-1}\cos [\mathcal{A}(x)\bar{\omega}_{i}
t]/\bar{\omega}_{i}\), where \(\mathcal{A}(x)=A {(B - x)}^2 -
1\). Hence, upon assuming \({\rho(\omega) = \alpha \omega^{2}}\), and
\({g(\omega) = \text{Const}}\), it appears possible to retrieve that
$K'(t)$ is proportional to $\delta(t)$ also in the relativistic case,
just like in the Galilean case discussed in
~\cite{zwanzig2001nonequilibrium}.  Unlike in the Galilean case,
however, the assumption \({g(\omega) = \text{Const}}\) appears
particularly strong and untenable in the relativistic case, because
assuming that the tagged particle can interact with the same coupling
strength with all the oscillator baths (irrespective of their
separation in space-time from the tagged particle) is at odds with the
principle of locality (which states that an object is directly
influenced only by its immediate surroundings).

Therefore, we expect that since \({g(\omega) \neq \text{Const}}\) is
imposed by the principle of locality in special relativity for a given
physical system, the fluctuation-dissipation relation associated with
our relativistic generalised Langevin equation~\eqref{eq:lang-relat}
must be non-Markovian, as discussed in the next subsection.

\subsection{Fluctuation-dissipation
  relations}\label{sec:relat-fluct-diss}


Here we shall demonstrate that the force \(\mathbf{F}'_p\) indeed has
characteristics of thermal noise, thus qualifying
\cref{eq:langevin-non-symmetric} as a Langevin equation.

Consider that the bath-particle system is subject to a large number of
computer simulations. In each such simulation, the heat bath initial
conditions are taken at random, such that the probability density
function (PDF) is
\begin{equation}
  f (\mathbf{x}, \dot{\mathbf{x}}, \mathbf{q}, \dot{\mathbf{q}}) \propto
  \exp \left[ - \frac{E} {k_{B}T}\right]\!\!,\label{eq:init-cond}
\end{equation}
where \( E \) is the total energy of the system for the microstate
\((\mathbf{x}, \dot{\mathbf{x}})\); i.e.~a
Maxwell-Boltzmann~\cite{zwanzig2001nonequilibrium} or J\"{u}ttner
distributed random variable~\cite{Juettner,DunkelPRL}. In the Galilean
case, the system would have been in thermal equilibrium with respect
to a frozen or constrained system coordinate \({\mathbf{x}(0)}\). Under these assumptions
one can prove the fluctuation-dissipation
theorem by direct evaluation of time averages~\cite{zwanzig2001nonequilibrium}.

However, special relativity makes this route somewhat more complex:
one has more terms, and evaluating any term with \(\bar{\xi}\)
dependence requires knowing the trajectory.  With reference to
\cref{eq:stochastic-force-relativistic}, the first two terms in the
sum on the r.h.s.~are identical to the terms that one has in the
non-relativistic case (c.f.~\cref{eq:noise-classical}) and it was
shown by Zwanzig that they give $\langle \mathbf{F}_p(t)\rangle=0$ for
the non-relativistic stochastic force~\cite{Zwanzig1973,Cui}. Here we
have an additional term, the last term in
\cref{eq:stochastic-force-relativistic} proportional to
\(\mathbf{x}(0)\).  In general, this term is non-trivial to evaluate
since it depends on the bath oscillator's trajectory through
$\bar{\xi_{i}}(s)$. However, we can show, by analytical integration,
that for the case where the trajectory is parameterised by
\cref{eq:xi-hypo} (where \(\xi = A s {(x-B)}^2\)), which was obtained
from the numerics in our simulations, the result of the integral over
\(s\) evaluated at \(s=0\) is simply \(\cos \bar{\omega}_{i}t\). Hence
in this case also the last term in
\cref{eq:stochastic-force-relativistic} averages to zero, and
\(\langle \mathbf{F}'_p(t)\rangle=0\) also for the relativistic
case. Also in the hypothetical case \(\xi \), one obtains sinusoidal
functions i.e.~a combination
\({C \sin \bar{\omega}_i (t +\frac{1}{4}) + D \cos \bar{\omega}_i (t
  +\frac{1}{4})}\), with \(C\) and \(D\) some \(t\)-independent
constants, which also leads to
\({\langle \mathbf{F}'_p (t)\rangle=0}\).  Therefore, based on the
numerical data that we have for \({\bar{\xi}(s)}\), we can conclude
that the term \({\mathbf{F}'_p(t)}\) has zero average and qualifies as
the stochastic term or the noise in our relativistic Langevin
equation. For most situations we expect that the integral over $s$ in
\cref{eq:stochastic-force-relativistic} evaluates to a sinusoidal
function of $\omega_{i} t$ upon integrating away the $s$ dependence
and evaluating at $s=0$.

Proceeding in a similar way, since the last term of
\cref{eq:stochastic-force-relativistic} for the numerically -evaluated
trajectory form given by \cref{eq:xi-hypo} (and presumably also for
other more complicated forms) leads to a function proportional to
\(\cos \bar{\omega}_{i}t\), evaluation of the time-correlation
function of \(\mathbf{F}'_p\) gives a product of sinusoidal functions
for all the terms present in
\cref{eq:stochastic-force-relativistic}. Following the same strategy
as in~\cite{zwanzig2001nonequilibrium} at page 23, and of~\cite{Cui},
the products of two sinusoidal functions of argument
\(\bar{\omega}_i t\) and \(\bar{\omega}_i t'\) respectively, by
applying the standard trigonometric identity for the product of
trigonometric functions, lead to a generic function \(K\) of argument
\((t-t')\),
\begin{equation}\label{eq:theorem}
  \langle \mathbf{F}_{p}(t) \mathbf{F}_{p} (t') \rangle  =m k_{B}T K(t-t').
\end{equation}
Since coordinate time is relative, the right hand side of the
fluctuation dissipation theorem must depend on the velocity of the
observer. If one could find a physically significant frame of
reference, however, proper time can be used and \(t\) can be replaced
with \(\tau\) in the above relation.

Analogy with the Galilean case would suggest that the product inside
the averaging brackets is an inner product of two 4-vectors, while the
right hand side of the Galilean fluctuation dissipation theorem is a
rank 2 tensor. Thus the product of two fluctuations must indeed also
be the outer product of the force 4-vectors.

Hypothetical deviations from \cref{eq:theorem} which may occur for
more non-trivial $s$-dependencies of $\xi$ are considered in
Appendix~\ref{sec:deviations}.

In more general and complex settings, for complex trajectory
dependence of \(\bar{\xi}(t, \mathbf{x})\), simple methods of
integration may not be sufficient to arrive at a FDT in closed
form. However, one could opt for a path-integral approach to the
problem along the general lines of Ref.~\cite{Lubensky}. That task is,
however, well-beyond the scope of the current study and is left for
future work, along with ascertaining whether the second additional
term in \cref{eq:guess-fdt} can be observed in certain conditions.

Another approach would be to consider the less extreme relativistic
behavior, i.e.~\({\bar{\xi} \rightarrow 0}\). As we have mentioned
previously, this case has been studied extensively and relations
analogous to the fluctuation dissipation theorem (also known as
Einstein-Sutherland relations), were obtained e.g.~in
Ref.~\cite{hang}. These completely ignore any effects that may be due
to \(O(\xi)\) terms, so, while a useful point of reference, they are
not totally useful for fully relativistic conditions.

\section{Conclusion}
In this largely expository paper, we have provided a first-principles
derivation, from a microscopic Caldeira-Leggett particle-bath
Lagrangian, of a full-fledged relativistic Langevin equation,
\cref{eq:langevin-non-symmetric}.  In some of its limits, this more
general relativistic equation recovers commonly
accepted~\cite{Debbasch,Quasirel-langev-eq,RappVanHees} extensions of
the Langevin equation to relativistic media and relativistic
weakly-interacting particles. By relaxing some of the commonly used
approximations, led to a full-fledged and more general representation
of stochastic processes in relativistic media valid for more strongly
relativistic conditions.

The new fully-relativistic generalised Langevin
equation~Eq. \eqref{eq:langevin-non-symmetric}, or in fully covariant form Eq. \eqref{eq:covariant-langevin}, contains a new force term $\mathbf{F}'_{r}$ which is trajectory-dependent and requires further investigation in future work. Also, the Fourier modes associated with the bath oscillators, are modified into a plane wave-like form which is naturally covariant, with a relativistic correction length-scale $\xi$ with is both time and trajectory dependent. Based on our numerical data, this dependence is an off-set parabola of the form $\xi(x,t)=At(x-B)^{2}$. The possible generality of this form has to be further ascertained in future numerical studies.

The new fully-relativistic Langevin equation may predict fundamentally
new physics. Firstly, there is some evidence that the universe is
chiral at the very largest scales\cite{Minami_2020}. Our equation
suggests a mechanism for parity violation due to dissipation, so
analysing the results of the aforementioned experiment may be of
utmost interest. The relativistic correction terms of the same order,
also predict the presence of a weak restoring force that tends to
bring the particle to its initial state/position, that may be
observable by the same methods by which Sakharov
oscillations~\cite{Sakharov}, and Baryon Acoustic Oscillations
(BAO)~\cite{BAO_1,BAO_2}, are observed in the CMB radiation. Also,
this effective force vanishes upon moving to speeds that are small
compared to the speed of light, and is therefore a genuine new effect
due to the interplay of relativistic motion and dissipation, which
otherwise vanishes in the non-relativistic limit.

We have also, in a way, reached the limit of what the Caldeira-Leggett
particle-bath models can tell us about the underlying dynamics of
relativistic media. One could account for more phenomena by
considering:
\begin{enumerate*}[label=(\roman*)] 
\item suitable incorporation of proper and co-ordinate times;
\item couplings that depend on the position of the tagged particle;
\item harmonic modes of other kinds (e.g.~magnetic confinement
  potentials that lead to pure harmonic oscillations, etc.)
\item pair production;
\item proper field-theoretic extensions of the Caldeira- Leggett
  Lagrangian model by means of path-integral formalism for
  state-dependent diffusion processes~\cite{Lubensky}.
\end{enumerate*}

Of particular interest are the questions of pair production and
field-theoretic extensions. By definition of rest energy, at
velocities where corrections that we have neglected may become
important, the energies of the medium and the particle are sufficient
to randomly produce pairs of charged particles. This extension may be
of interest both for examining super-heated exotic objects, as well as
the early universe.  Further important applications are in the context
of nuclear physics, where so far only the Galilean generalised
Langevin equation has been used to describe fission
processes~\cite{Kolomietz2}, while clearly relativistic corrections
may be important for the fission of hot nuclei. Finally, the results presented here could be a first step for a more extensive and detail analysis of thermodynamic aspects of
relativistic systems \cite{Tolman}, including heat distribution \cite{Paraguassu}, and fluctuation theorems \cite{Pal_2020}.

In future work, one could also attempt to produce the
General-Relativistic (GR) extension of the Langevin equation by the
same methodology presented in this paper. 
Finally, it will be interesting to compare results from the above derivations with certain limits of Langevin equations
obtained using the Schwinger-Keldysh formalism within AdS/CFT approaches \cite{Teaney,Mueller}.

\section*{Acknowledgments}
Matteo Baggioli and Bingyu Cui are gratefully acknowledged for
discussions and assistance in the numerical calculations.

W. J. Handley is acknowledged in assistance in identifying suitable
cosmological applications of the results.

A. Zaccone gratefully acknowledges financial support from the US Army
Research Office through grant \textnumero~W911NF--19--2--0055.

\appendix

\section{The dynamical coupling model in detail}\label{sec:time-lag-coupling}

To answer this question we must more deeply elaborate on the origin of
the modes. The real underlying model is that of hard collision
potentials between the tagged particle and the constituent particles
of the heat bath. There is a different probability of each collision
between each species of particle, and multiple possible modes of
interaction that we all ignore. Every potential that would lead to
both the tagged particle and the heat bath confined to the region
that we discussed, entails describing both the generalised
\(\mathbf{q}_{i}\) and the dynamical \(\mathbf{x}\) as displacements
from a common equilibrium. Thus \(\dot{\mathbf{x}}\) is always
interpreted as a relative velocity with respect to the bulk velocity
of the heat bath.

In the relativistic case, the collisions are highly local, so one
might think that they can safely assume \(t_{i} \approx t\). However,
the interaction with the heat bath modes entails the interaction of
the particle that was hit by the tagged particle with the rest of
the heat bath, and generate a bosonic excitation. This interaction is
not local in general. Instead, to avoid substantial complications in
the Euler-Lagrange equations, we shall note that the difference
between \(\mathbf{q}_{i}(t_{i})\) and \(\mathbf{q}_{i}(t)\) depends on
the extent of the heat bath that we consider: the smaller the quantity
of particles participating, the smaller the effective distance between
\(t\) and \(t_{i}\). Thus we restrict ourselves to interactions
between local modes, where
\(\mathbf{q}_{i}(t_{i}) \approx \mathbf{q}_{i}(t)\). This has the
added benefit of making virtually any of the initial conditions that
were deemed acceptable in the Galilean case, to be useful in the
Lorentz covariant case (cfr.~\ref{sec:integr-const-init}).

\section{On the limit of \( g \mathbf{x} \leq 2 \) }\label{sec:gx}

The separation between the modes in the units of \(c=1\) (and with
\(m = m_{i} = 1\) for convenience), only causes problems in
\cref{eq:eom-rel-q} \emph{if an only if} the right hand side were to
become negative, which occurs, provided \(\omega = 1\) when
\(g \mathbf{x} > 2\). Because \(\mathbf{x}\) only enters the equations
of motion as part of \(g \mathbf{x}\), we expect that the solution
also, only depends on the combination \(g \mathbf{x}\). So with a
stronger coupling, we expect a narrower range of tagged particle
displacement to not violate the assumptions. And vice versa, weaker
coupling allows for broader variations in \(\mathbf{x}\), as seen in
\cref{fig:xi-contour}.

Firstly, we see that the constraints set by causality can be
implemented by constraining the range of \(\mathbf{x}\) for a fixed
\(g\). Secondly, the units we are using are that in which the
characteristic oscillation frequency is unity. In the LHC, the
relevant frequency would the plasma frequency of the electrons in the
cloud. By taking the average electron number density~\cite{plasma}
\(n_{e} \approx \SI{.e7}{\per \cubic \centi \metre} = \SI{.e13}{\per
  \cubic \metre}\) we get a plasma frequency
\begin{equation}
  \omega_{pe}
  =
  \sqrt{\frac{n_{e}e^{2}}{\epsilon_{0}m_{e}} + 3k_{B}v_{th, e}^{2}}
  \approx \SI{1.7}{\kHz}, \label{eq:plasma}
\end{equation}
thus, \(\mathbf{x} = 1\) translates to approximately
\(\SI{150}{\km}\), six times the circumference of the LHC.\@For a star
like e.g.~Sol, the plasma frequency is much higher, and the size of
the core is much closer to \(\SI{1.e5}{\km}\)~\cite{solarRadius}. Thus
one does not expect many relativistic effects at the LHC, but does at
the core of the sun.

In either case, the distance at which the boundary conditions can be
considered constant are far beyond the distance at which
\( g \mathbf{x} > 2 \) could cause issues.

It must also be noted, that \(g=2\) is an unrealistic coupling
strength. The Caldeira-Leggett Hamiltonian corresponding to
\cref{eq:c-l-classical} and \cref{eq:starting-lagrangian-rel}, is
responsible for the heat capacity of the medium. If for the case of
\(\omega=1\), \(g=2\) were the true value, the heat capacity of the
medium would be strongly affected by the presence of the tagged
particle. This is not what we observe in the Galilean case, thus
ruling out such a possibility for the Lorentz covariant case. The
reason why we us values of \(g=2\) is purely illustrative, as the true
solution to the Euler-Lagrange equations can only depend on
\(g \mathbf{x}\) and the plots are easier to read for reasonable
values of \(\mathbf{x}\).

\section{Equivalence of time-dependent frequency and phase}\label{sec:tdfrequency}
Here we demonstrate that for an arbitrary phase relationship, the
time-dependence of the frequency can be without loss of generality be
shifted onto a time dependent phase.

Let \( \phi(t) \) and \( \omega(t) \) be some time-dependent
functions. We can always re-write the phase of the form
\begin{equation}
  \omega(t) \left( t - \phi(t)\right) \label{eq:phase}
\end{equation}
as
\begin{equation}
  \Omega \left( t - \Phi(t) \right) = \omega(t) \left( t - \phi(t)\right)\label{eq:also-phase}
\end{equation}
where \( \Omega = \text{Const.} \) To do so, recognise that the change
of variables has two degrees of freedom, \( \Omega \) and \(\Phi\),
but only one constraint --- \cref{eq:also-phase}. Thus, by setting
\begin{equation}
  \Phi(t) \equiv - \omega(t)\left(t - \phi(t)\right) - \Omega t
\end{equation}
one can easily verify that \cref{eq:also-phase} holds as an identity.
Thus one can \emph{always} choose to deal with a time-dependent phase
in lieu of a time-dependent frequency, where, as in the following
appendix, the integration is greatly simplified by not considering
time-dependent frequencies in addition to an already implicitly
time-dependent phase.

  \section{Integrating the tagged particle and heat bath equations of
    motion}\label{sec:integr-tagg-part}

  Let's derive \cref{eq:bath-motion-relat-int}. For simplicity easier,
  we have shifted the \(\mathbf{x}\) dependence of \(\bar{\omega}\),
  onto \(\bar{\xi}\).As a result we can treat \(\bar{\omega}\) as a
  constant during integration, and \(\bar{\xi}\) picks up an
  additional explicit co-ordinate time dependence, which does not
  affect the integration as opposed to
  \(\bar{\xi} = \bar{\xi}(\mathbf{x}(t))\), because for all intents
  and purposes the two functions are unknown and depend on time.
\begin{widetext}
  We start with the Newton-Leibniz formula.
  \begin{align}
    \label{eq:by-parts-2}
    \begin{split}
      \int_{0}^{t} \gamma(\mathbf{\dot{x}}(s)) \mathbf{x}(s) \frac{\sin
        \bar{\omega}(t - \frac{\bar{\xi}(t)}{c} - s)}{\bar{\omega}} ds =
      &\frac{\mathbf{x}(t)}{\bar{\omega}} \substitute{\int
        \gamma(\mathbf{\dot{x}}(s))\sin \bar{\omega} \left(t -
          \frac{\bar{\xi}(s)}{c} - s\right) ds }{s=t}\\
      &- \frac{\mathbf{x}(0)}{\bar{\omega}} \substitute{\int
        \gamma(\mathbf{\dot{x}}(s))\sin \bar{\omega}\left(t -
          \frac{\bar{\xi}(s)}{c} - s\right) ds }{s=0} \\ & - \int_{0}^{t} \gamma
      (\mathbf{\dot{x}}(s)) \mathbf{\dot{x}}(s) \left\{
        \gamma^{-1}(\mathbf{\dot{x}}(s)) \int\gamma (\mathbf{\dot{x}}(s))
        \frac{\sin \bar{\omega}(t - \frac{\bar{\xi}(s)}{c} -s)}{\bar{\omega}} \,
        ds\right\} ds.
    \end{split}
  \end{align}
  The main difficulty arises due to the time dependence in
  \(\bar{\xi}(t, \mathbf{x}(s)) = \bar{\xi}(s)\).

  Here we must only obtain a combination that enters the force given
  by the Euler-Lagrange equations:
  \({\mathbf{\dot{q}} - g \mathbf{\dot{x}}/ \bar{\omega}^{2}}\),
  consequently, in lieu of prolonged algebraic manipulations, we will
  merely extract a factor of \(\mathbf{x}(t)\).
  \begin{align}
    \label{eq:parts-pre-final}
    \begin{split}
      \frac{\mathbf{x}(t)}{\bar{\omega}}
      \substitute{
        \!\int \gamma(\mathbf{\dot{x}}(s))\!\sin \bar{\omega}\!
        \left(\!t\!-\!s\!-\!\frac{\bar{\xi}(s)}{c} \right)\, ds
      }
      {t}
      =
      \frac{\mathbf{x}(t)}{\bar{\omega}}
      \left\{ 1 +  \substitute{
          \!\int \gamma(\mathbf{\dot{x}}(s))\!\sin \bar{\omega}\!
          \left(\!t\!-\!s\!-\!\frac{\bar{\xi}(s)}{c} \right)\, ds
        }
        {t} - 1 \right\}.
    \end{split}
  \end{align}
  Hence, collecting terms we get, for the $i$-th oscillator:
  \begin{align}\label{eq:bath-motion-relat-int-app}
    \begin{split}
      \mathbf{q}_{i}(t)\!-\!\frac{g_i\mathbf{x}(t)}{{(\bar{\omega}_i)}^2}
      = &\left[ \mathbf{q}_{i}(0) - \frac{ g_i}{\bar{\omega}_i^2}
        \mathbf{x}(0) \right] \cos \bar{\omega}_i \left( t -
        \frac{\bar{\xi}_{i}(t)}{c} \right)+ \gamma(\mathbf{\dot{q}}_i (0))\mathbf{\dot{q}}_i (0) \frac{\sin
        \bar{\omega}_i \left( t -\frac{\bar{\xi}_{i}(t)}{c} \right)}
      {\bar{\omega}_i} \\
      &- g_i\frac{\mathbf{x}(t)}{\bar{\omega}_{i}^{2}}
      \substitute{\int \bar{\omega}_{i} \gamma(\mathbf{\dot{x}}(s)) \sin \bar{\omega}_{i}\left(t - \frac{\bar{\xi}(s)}{c} -s\right) ds - 1}{t} \\
      &+ \int_{0}^{t} \gamma
      (\mathbf{\dot{x}}(s)) \mathbf{\dot{x}}(s) \left\{
        \gamma^{-1}(\mathbf{\dot{x}}(s)) \int\gamma (\mathbf{\dot{x}}(s))
        \frac{\sin \bar{\omega}_{i}(t - \frac{\bar{\xi}(s)}{c} -s)}{\bar{\omega}_{i}} \,
        ds\right\} ds\\
      &+  g_i \frac{\mathbf{x}(0)}{\bar{\omega}_{i}^{2}} \substitute{\cos
        \bar{\omega}_i \left( t - \frac{\bar{\xi}_{i}(t)}{c} \right) + \int
        \gamma(\mathbf{\dot{x}}(s)) \bar{\omega}_{i} \sin \bar{\omega}_{i}(t -
        \frac{\bar{\xi}(s)}{c} -s) ds}{0}. \\
    \end{split}
  \end{align}
\end{widetext}

\section{Hypothetical deviations from \cref{eq:theorem}}\label{sec:deviations}
We should also consider the hypothetical case where the last term on
the r.h.s.~of \cref{eq:stochastic-force-relativistic} does not simply
lead to a sinusoidal function of $\bar{\omega_i}t$, in which case we
are left with an additional term in the FDT, i.e.~in the r.h.s.~of
\cref{eq:theorem}.  While the equations of motion
\cref{eq:langevin-non-symmetric} and \cref{eq:langevin-classical} are
different, the latter is the low velocity limit of the former. It was shown \cite{langevin_force}  that the Galilean Langevin
equation with an additional time-dependent external ``AC'' potential,
supports the fluctuation-dissipation theorem plus a correction term
arising from the AC field. By applying
to~\eqref{eq:stochastic-force-relativistic} similar manipulations as
~\cite{langevin_force} in dealing with the additional term (the origin
of which in our case is entirely relativistic), in the reference frame
of the tagged particle one could expect
\begin{equation}
    \left\langle \mathbf{F}'_p (\tau) \mathbf{F}'_p(\tau')\right\rangle \approx m k_{B} T K(\tau - \tau') + \left\langle \mathbf{F}'_r(\tau) \mathbf{F}'_r (\tau') \right\rangle. \label{eq:guess-fdt}
\end{equation}
The first term on the r.h.s.~is the standard FDT as derived by Zwanzig
using the same procedure for the non-relativistic Langevin
equation~\cite{Zwanzig1973,zwanzig2001nonequilibrium} and as we found
above for the case of our simple, numerics-informed, relativistic
calculations.  We cannot \emph{a priori} exclude that the second term
on the r.h.s.~of the above \cref{eq:guess-fdt} will appear for more
complex trajectories in the relativistic Langevin equation; however,
it correctly vanishes in the limit of non-relativistic velocities,
where \(\mathbf{F}'_r\) vanishes, as discussed above.

Although our numerical calculations support a rather standard
non-Markovian FDT akin to \cref{eq:theorem}, further investigation
into physical instances where correction terms similar to those in
\cref{eq:guess-fdt} is necessary.

\bibliography{bibliography}

\end{document}